\documentclass[%
 reprint,
superscriptaddress,
 amsmath,amssymb,
 aps,
prb,
]{revtex4-2}

\usepackage{graphicx}% Include figure files
\usepackage{dcolumn}% Align table columns on decimal point
\usepackage{bm}% bold math
\usepackage{hyperref}
\usepackage{soul}
\newcommand{\KYS} {KYbSe$_2$}
\newcommand{\TN} {$T_{\text{N}}$}
\newcommand{\Tsample} {$T_{\text{sample}}(H)$}

\begin{document}

%\preprint{APS/123-QED}

\title{Magnetic field-temperature phase diagram of spin-1/2 triangular lattice antiferromagnet KYbSe$_{2}$}

\author{Sangyun Lee}
 \thanks{These authors contributed equally to this work.}
 \affiliation{Los Alamos National Laboratory, Los Alamos, New Mexico 87545, USA.}

\author{Andrew J. Woods}%
 \thanks{These authors contributed equally to this work.}
\affiliation{Los Alamos National Laboratory, Los Alamos, New Mexico 87545, USA.}

\author{Minseong Lee}
\affiliation{Los Alamos National Laboratory, Los Alamos, New Mexico 87545, USA.}

\author{Shengzhi Zhang}
\affiliation{Los Alamos National Laboratory, Los Alamos, New Mexico 87545, USA.}

\author{Eun Sang Choi}
\affiliation{National High Magnetic Field Laboratory, Florida State University, Tallahassee, Florida 32310-3706,  USA.}

\author{A. O. Scheie}%
% \affiliation{Neutron scattering division, Oak ridge national laboratory, Oak ridge, TN 37831, USA.}
\affiliation{Los Alamos National Laboratory, Los Alamos, New Mexico 87545, USA.}

\author{D. A. Tennant}
\affiliation{Neutron Scattering Division, Oak Ridge National Laboratory, Oak Ridge, TN 37831, USA.}

\author{J. Xing}
\affiliation{Materials Science and Technology Division, Oak Ridge National Laboratory, Oak Ridge, Tennessee 37831, USA.}

\author{A. S. Sefat}
\affiliation{Materials Science and Technology Division, Oak Ridge National Laboratory, Oak Ridge, Tennessee 37831, USA.}

\author{R. Movshovich}%
\email{roman@lanl.gov}
\affiliation{Los Alamos National Laboratory, Los Alamos, New Mexico 87545, USA.}

\date{\today}% It is always \today, today,

\begin{abstract}
A quantum spin liquid (QSL) is a state of matter characterized by fractionalized quasiparticle excitations, quantum entanglement, and a lack of long-range magnetic order. However, QSLs have evaded definitive experimental observation. Several Yb$^{3+}$-based triangular lattice antiferromagnets with effective $S$ = $\frac{1}{2}$ have been suggested to stabilize the QSL state as the ground state. Here, we build a comprehensive magnetic temperature phase diagram of a high-quality single crystalline {\KYS } via heat capacity and magnetocaloric effect down to 30~mK with magnetic field applied along the $a$-axis. At zero magnetic field, we observe the magnetic long-range order at {\TN } = 0.29~K entering 120~degrees ordered state in heat capacity, consistent with neutron scattering studies. Analysis of the low-temperature ($T$) specific heat ($C$) at zero magnetic field indicates linear $T$-dependence of $C/T$ and a broad hump of $C/T$ in the proximate QSL region above {\TN}. By applying magnetic field, we observe the up-up-down phase with 1/3 magnetization plateau and oblique phases, in addition to two new phases. These observations strongly indicate that while {\KYS} closely exhibits characteristics resembling an ideal triangular lattice, deviations may exist, such as the effect of the next-nearest-neighbor exchange interaction, calling for careful consideration for spin Hamiltonian modeling. Further investigations into tuning parameters, such as chemical pressure, could potentially induce an intriguing QSL phase in the material.
\end{abstract}

\maketitle

%\tableofcontents

\section{\label{sec:Introduction}Introduction}
Quantum spin liquids (QSLs) are a proposed state of matter characterized by fractionalized quasiparticle excitations, quantum entanglement, and a lack of long-range magnetic order \cite{balents2010spin,savary2016quantum,broholm2020quantum,zhou2017quantum}. This elusive phase of matter has captured great attention from researchers because quantum entanglement can be utilized to realize fault-tolerant quantum computing \cite{nayak2008non,kitaev2003fault}. Researchers have been actively searching for the QSL phase in materials with magnetic frustrations. Among the simplest geometrically frustrated systems are the triangular lattice antiferromagnets (TLAFs) \cite{anderson1973resonating,moessner2006geometrical}. Recent reports have suggested the possibility of TLAFs hosting a gapless U(1) Dirac QSL phase \cite{hu2019dirac} or a chiral spin liquid with spinon Fermi surfaces \cite{gong2019chiral} at zero magnetic field when the ratio of the nearest-neighbor ($J_{1}$) and next nearest-neighbor ($J_{2}$) exchange interactions is close to 0.06 ($J_{2}/J_{1}$ = 0.06) \cite{manuel1999magnetic, kaneko2014gapless,iqbal2016spin,li2015quasiclassical, hu2015competing, saadatmand2016symmetry, wietek2017chiral, gong2017global, bauer2017schwinger}.

The strong magnetic frustration in theoretical TLAFs induces an interesting magnetization process under magnetic field: a 120-degree ordered state at zero magnetic field (if a long-range ordering exists at zero magnetic field) and an up-up-down (UUD) spin structure with a 1/3 magnetization plateau within a finite range of the magnetic field \cite{nishimori1986magnetization, chubukov1991quantum, susuki2013magnetization}. The UUD phase is a fluctuation-driven magnetic phase whose spin structure differs from the classical ground state, through the order-by-disorder mechanism \cite{villain1980order}. The field range of the UUD phase decreases as the temperature decreases for large spin TLAFs, as the fluctuation is driven by thermal fluctuations \cite{seabra2011phase,lee2014magnetic}. Conversely, for small spin TLAFs, the UUD phase persists down to zero temperature due to the dominance of quantum fluctuations \cite{chubukov1991quantum,quirion2015magnetic, lee2014series, zhou2012successive}. As the magnetic field further increases, an oblique spin structure is also stabilized before the magnetization reaches saturation.

Recently, delafossite materials with the $ARX_{2}$ structure have been investigated as promising candidates for quantum spin liquids (QSLs) \cite{schmidt2021yb, xing2019crystal, zangeneh2019single, scheie2020crystal, bordelon2021frustrated}. These materials feature two-dimensional triangular lattices of $R^{3+}$ ions, where $A$ represents alkali metals (Li, Na, K, Rb, Cs, Tl, Ag, and Cu), $R$ represents rare-earth elements, and $X$ represents chalcogen elements (O, S, and Se). The crystal structure is illustrated in Fig.~\ref{fig:Fig1} (a). Rare-earth elements form a triangular lattice, allowing for tuning the distance between nearest neighbor rare earth ions. Alkali metals modulate interlayer interactions, while chalcogen elements affect intralayer interactions \cite{schmidt2021yb, xing2019crystal, zangeneh2019single, scheie2020crystal}. The diverse chemistry that controls the physical properties gives rise to a series of compounds that allow exploration of various intriguing ground states, including the possibility of realizing QSL phases \cite{schmidt2021yb, xing2019crystal, zangeneh2019single, scheie2020crystal, bordelon2021frustrated, salke2014phase, boyraz2022effect}. 

Significant studies have been conducted on Yb$^{3+}$-based materials due to the $S$ = 1/2 state, which maximizes the quantum effect \cite{sichelschmidt2020effective, wu2019tomonaga}. Among the $A$YbSe$_{2}$ family of delafossite materials, {\KYS } has been considered a candidate for QSL. However, at temperatures below 0.29~K, magnetic order has been reported based on specific heat and neutron scattering experiments without a clear signature of gap opening \cite{scheie2023proximate, scheie2022non}. When a magnetic field is applied in the $ab$-plane, field-induced long-range magnetic orders are found at the 1/3 saturation magnetization plateau, most likely identified as UUD \cite{xing2021synthesis, yamamoto2014quantum}. 

Although {\KYS } exhibits magnetic order at low temperatures, the results indicate that it is a proximate QSL, as suggested by the $J_{2}/J_{1}$ ratio estimated using specific heat and inelastic neutron scattering data~\cite{scheie2023proximate, scheie2022non}. The $J_{2}/J_{1}$ ratio extracted from neutron scattering and heat capacity data (0.047) is very close to that (0.06) of the QSL quantum phase transition on the triangular lattice Heisenberg antiferromagnet. Therefore, {\KYS } serves as a significant milestone towards realizing a QSL, with the potential for fine-tuning through methods such as chemical doping and pressure. 

Here, we have studied the thermodynamic properties of {\KYS } using heat capacity and the magnetocaloric effect (MCE) and have constructed a comprehensive in-plane magnetic field vs. temperature ($H-T$) phase diagram. We have observed a single ordering transition at 0.29~K at zero magnetic field. Through the convergence of data from both heat capacity and MCE analyses, consistent magnetic phase boundaries have been obtained. The nature of the magnetic phase transitions closely aligns with second-order phase transitions, suggesting a negligible magnetic anisotropy within the plane. Furthermore, our phase diagram exposes five distinct magnetic phases, encompassing a 120-degree ordered phase, a UUD phase, and an oblique phase.
%and a $T$-linear feature in the specific heat divided by temperature ($C/T$), which is consistent with the signature of a proximate gapless QSL with Dirac cone-like dispersion. Through analysis of the heat capacity data, we have extracted the magnetic moment of each Yb$^{3+}$ ion, the energy gap, and the residual specific heat (the Sommerfeld coefficient) at zero magnetic field.
In light of our comprehensive findings, it is apparent that the compound under investigation exhibits characteristics closely resembling those of an ideal TLAF. However, the revelation of two additional phases hints at the possible existence of supplementary magnetic interactions.
%our phase diagram reveals five different magnetic phases, including a 120-degree ordered phase, UUD phase and oblique phase. while it is evident that the compound under investigation closely approaches the characteristics of an ideal TLAF, the two additional phases may indicates the existence of additional magnetic interactions.

\section{\label{sec:Exp_Details}Experimental Details}

{\KYS } single crystals were synthesized using a KCl salt flux method, and the detailed synthesis methods are described elsewhere \cite{xing2021synthesis}. All specific heat and MCE measurements presented in this study were conducted in an Oxford Instruments dilution refrigerator with a temperature range of 30~mK to 1.6~K and an applied magnetic field of up to 11~T. Specific heat values were obtained using a quasiadiabatic method. The MCE was also measured using the same experimental setup. The bath temperature was fixed, and the magnetic field was swept at rates between 0.09 and 0.19~T/min while monitoring the sample temperature.

For the measurements, a heater was mounted on one side of a sapphire stage, while a large single crystal sample of {\KYS } (1.19~mg) was mounted on the other side of the sapphire stage using GE varnish. The sample was positioned with the $a$-axis parallel to the direction of the magnetic field ($H \parallel a$). A ruthenium oxide resistance thermometer was affixed to the top of the sample, and a weak thermal link to the temperature bath was established by directly applying GE varnish to the sample. Due to the negligible heat capacity of the thermometer compared to the sample, the heat transfer between the thermometer and the sample was minimal.

\section{\label{sec:res} Results}

\begin{figure}
	\centering
    \includegraphics[width=1\columnwidth]{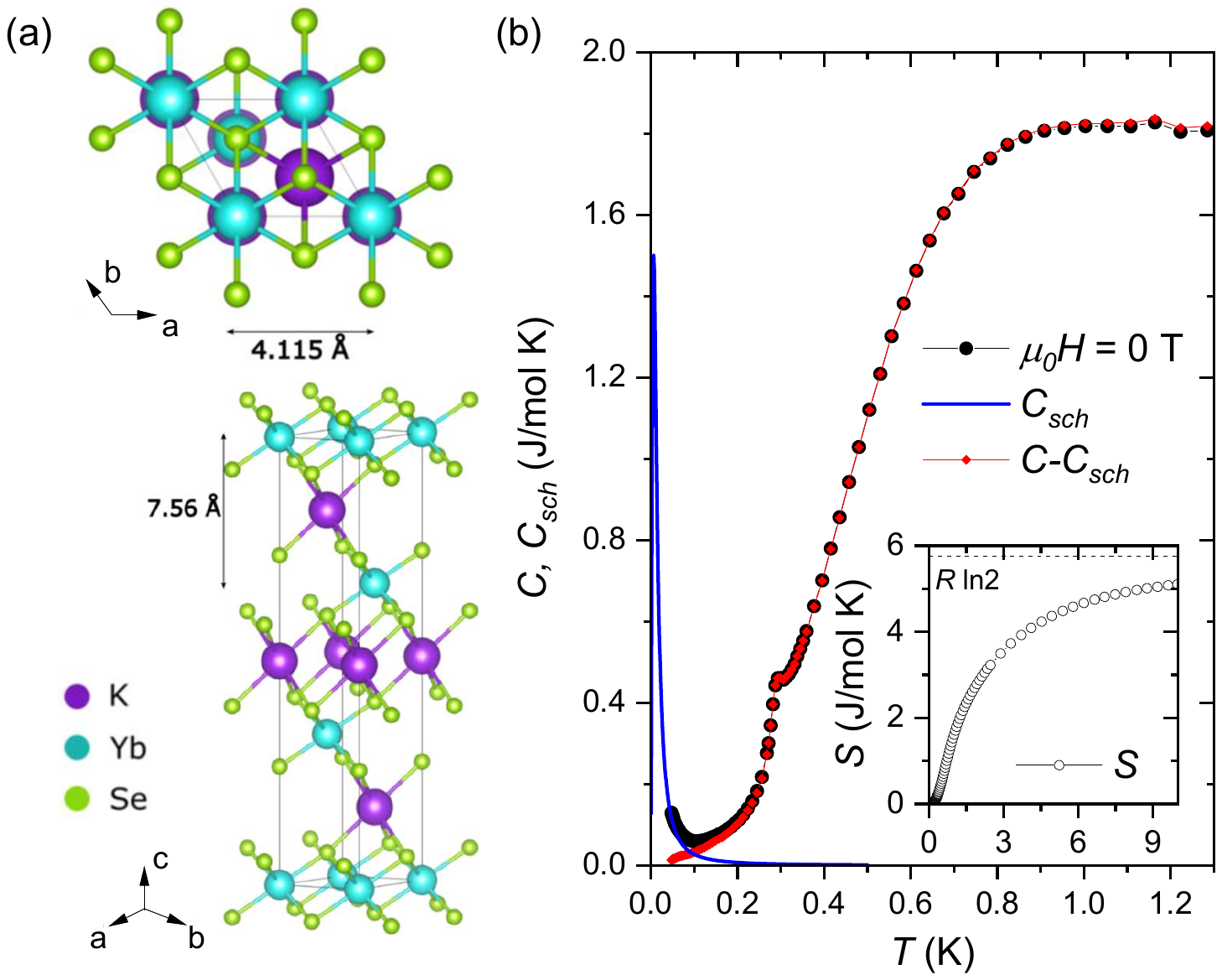}
\caption{\label{fig:Fig1}The crystal structure of KYbSe$_2$ (a) and specific heat, $C$ as a function of temperature, $T$ of single crystal KYbSe$_2$ in zero applied magnetic field (b). The black solid circles are the total specific heat, the blue line is a nuclear Schottky fit, $C_{sch}$ to the low-$T$ upturn, and the red solid circles are the data with the nuclear Schottky contribution subtracted. The inset shows the magnetic entropy, $S$ as a function of temperature.}
\end{figure}

In Fig.~\ref{fig:Fig1} (b), the zero-field heat capacity of the single crystal {\KYS } is shown as a function of temperature. At $T$ = 0.29 K, a small peak is clearly observed, which we attribute to a transition into a 120-degree ordered state, consistent with the neutron scattering results \cite{scheie2023proximate}. Below 0.1~K, the specific heat ($C$) starts to increase with decreasing temperature due to the nuclear Schottky anomaly of the Yb$^{3+}$ ions. The contribution of the nuclear Schottky anomaly is represented by the solid blue line ($C_{\text{sch}}$). Here, $C_{\text{sch}}$ = $dE/dT$, where $E = N \Sigma_{i=0}^n \Delta_i exp(-\Delta_i/k_BT)/\Sigma_{i=0}^n exp(-\Delta_i/k_BT)$. $n$ is the number of split the ground state separated by $\Delta_i$~\cite{Gopalsch}. 
so that 
\begin{equation}
C_{\text{sch}} = \frac{1}{Z k_B T^2} \Big[
\sum_{i=0}^n \Delta_i^2 e^{\frac{-\Delta_i}{k_B T}} -
\frac{1}{Z} \big( \sum_{i=0}^n \Delta_i e^{\frac{-\Delta_i}{k_B T}} \big)^2
\Big].
\label{eq:NuclearSchottky}
\end{equation}
The Hamiltonian for the nuclear hyperfine level splitting $\Delta_i$ is given by
\begin{equation}
\mathcal{H} = a \langle J_z \rangle I_z + P\big(I_z^2 - \frac{1}{3}I(I+1)\big)
\label{eq:hyperfineHamiltonian}
\end{equation}
%{\bf NEED TO WRITE FORMULA HERE} 
where $\langle J_z \rangle$ is the expectation value of the electronic magnetic moment, and $I$ and $I_z$ refers to the nuclear magnetic moment. ($a$ is the dipolar hyperfine constant, and $P$ is the quadripolar coupling constants which we take from Ref. \cite{Bleaney1963}). If we neglect the quadrupolar term, the upturn in the nuclear Schottky anomaly can be used to put an upper bound on the root-mean-squared $\langle J_z \rangle$ static electronic magnetic moment \cite{scheie2019exotic}. 

After subtracting the nuclear Schottky tail, the magnetic entropy ($S$) is estimated by integrating $C/T$ with respect to temperature, as shown in the inset of Fig.~\ref{fig:Fig1} (b). The calculated $S$ recovers approximately 89\% of $R$$\ln 2$ up to 10~K. The slow recovery is attributed to the magnetic frustration that suppresses the magnetic ordering temperature compared to the interaction energy scale. One important observation is that only one-phase transition at zero magnetic field was observed. This implies that the system is nearly isotropic \cite{ishii2011successive} and the spins are not strongly affected by thermal fluctuations \cite{seabra2011phase}.

\begin{figure}
	\centering
    \includegraphics[width=1\columnwidth]{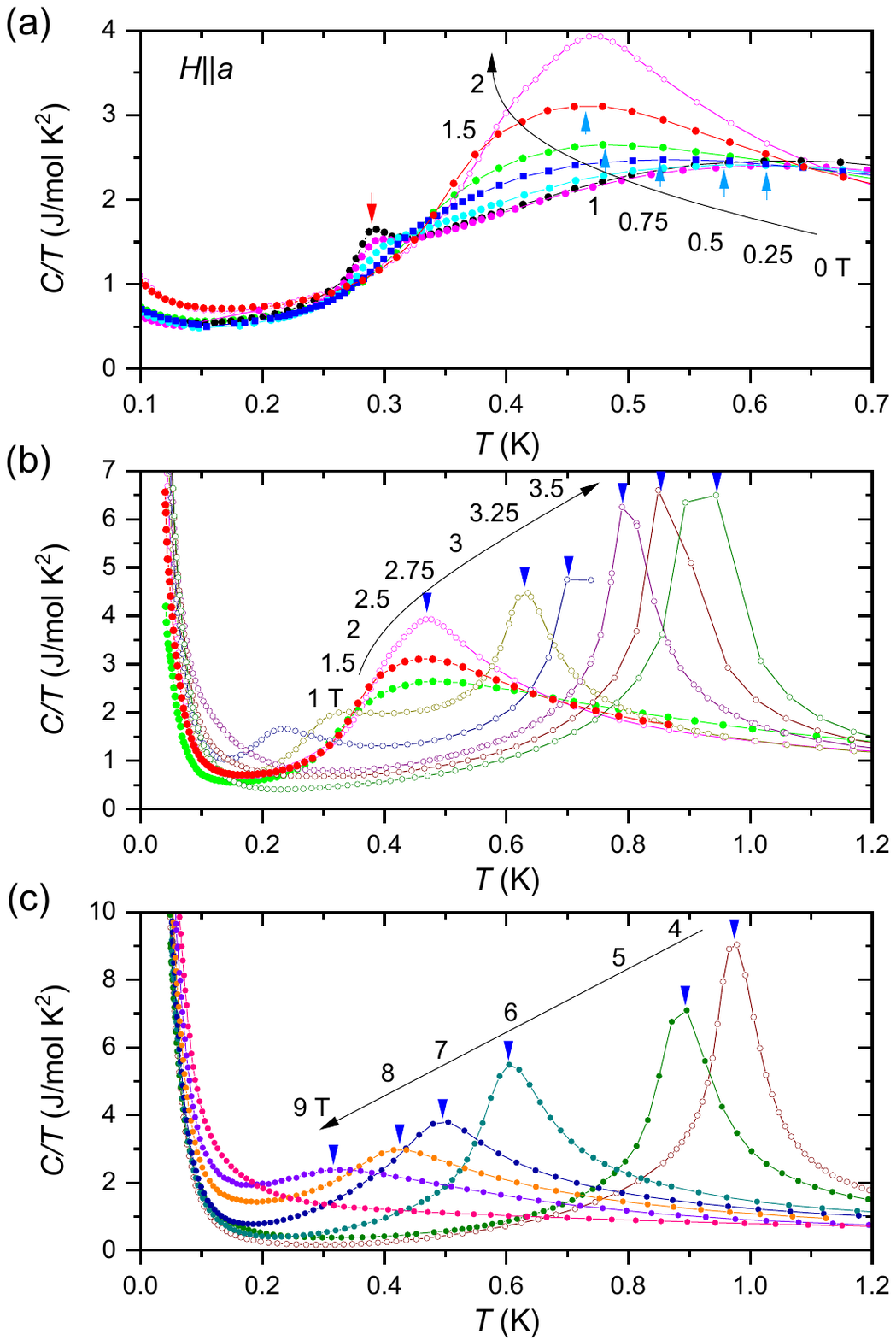}
\caption{\label{fig:HC1}Specific heat divided by temperature, $C/T$ as a function of temperature of single crystal KYbSe$_2$ in diﬀerent applied magnetic fields (a) Field range 0$-$2~T, the red arrow indicates the ordering transition, the blue arrows show the development of a broad hump. (b) Field range 0$-$3.5~T, the blue arrows show the development of a strong lambda-like anomaly. (c) Field range 4$-$10~T, the blue arrows indicate the continued development of the lambda-like transition.}
\end{figure}

We next measured $C/T$ as a function of temperature under various constant applied magnetic fields along $a$-axis, and the results are shown in Fig.~\ref{fig:HC1}. In Fig.~\ref{fig:HC1} (a), the ordering transition is indicated by a red arrow. As the magnetic field is increased, the transition weakens and becomes difficult to identify when $H > 1$ T while the height of the broad humps indicated with the light blue arrows increases. However, when $H \geq 2$ T, a prominent lambda-like peak marked with the blue arrows begins to develop, as shown in Fig.~\ref{fig:HC1} (b) and (c). This peak increases in both intensity and temperature, reaching 0.94~K at 4~T, indicating a rapid increase in entropy across the transition. As the field is further increased above 4~T, the lambda-like peak is suppressed, as shown in Fig.~\ref{fig:HC1} (c).

To detect phase transitions as a function of magnetic field, we measured $C/T$ while scanning the magnetic field at a constant temperature, as shown in Fig.~\ref{fig:HC2}. For clarity, $C/T$ is vertically shifted by $\alpha$ with the temperature evolution, and each constant $\alpha$ is in parentheses beside the legend.  We identify a series of magnetic phase transitions, and the transition fields are indicated by arrows in Fig.~\ref{fig:HC2}. Our data above 0.4 K match well the previously studied field-induced magnetic phase transitions \cite{xing2021synthesis}. However, $C/T$ below 0.4 K reveals additional new features at low fields below 2~T (denoted by the light red arrows) and high fields around 8~T (denoted by the light blue arrows), whose features are weak but clear. 

Next we turn to the MCE, which measures the change in the sample temperature while sweeping magnetic field {\Tsample }. The MCE can be measured in different conditions: adiabatic, quasiadiabatic, equilibrium, and isothermal conditions. In the adiabatic limit, where the time constant for the temperature of the sample to relax to the bath is much longer than the magnetic field sweep rate, the magnetic field up-sweeps and down-sweeps {\Tsample } are identical, i.e., it is reversible \cite{zapf2014bose, lee2023field}. A first-order phase transition with latent heat can make it irreversible. On the other hand, when the time constant for relaxation is much shorter than the field sweep rate, the equilibrium limit is obtained and the {\Tsample } displays sign-changing of slope near the phase boundaries on magnetic field up-sweeps and down-sweeps \cite{zapf2014bose}. Fig.~\ref{fig:mce} illustrates {\Tsample } for up (black) and down (red) sweeps of the magnetic field for various initial temperatures in zero magnetic field. The sign-change in slope around the phase boundaries indicates that our measurements were taken under close equilibrium conditions. This condition was anticipated due to the maximum field sweep rate of only 0.19~T/min with a fairly good thermal link to the bath as described in \autoref{sec:Exp_Details}. Starting from the initial temperature of 1.05~K in the paramagnetic state, {\Tsample } displays a broad hump around 0.7~T and progressively decreases as the field strength increases from 3~T to 5~T. During the down-sweep, {\Tsample } exhibits a gradual rise while the field decreases from 8~T to 5~T, followed by a decrease between 5~T and 3~T, eventually converging to approximately 1.0~K at zero magnetic field. {\Tsample } starting at 0.85~K and 0.63~K displays qualitatively analogous curves but with more pronounced features. Notably, {\Tsample } curves starting below 0.4~K exhibit additional features. A new peak feature arises around 2~T, shifting towards higher fields with decreasing temperature. Furthermore, a much clearer sign-changing phenomenon emerges around 3~T and 5~T. The broad humps at low field around 1~T grow more substantial as the initial temperature decreases.
\section{\label{sec:dis}Discussion}

We have developed a magnetic field versus low-temperature phase diagram by plotting the critical temperatures and magnetic fields from the specific heat and MCE data from \autoref{sec:res}. The resulting diagram is presented in Fig.~\ref{fig:PD}. Although some data points contain uncertainty that introduces a sizable error bar, distinct patterns within the phase diagram are evident, and five distinct phases are clearly delineated. For the specific heat data, we determine the temperature and magnetic field values at which the temperature (magnetic) derivative of $C/T$ becomes zero for the phase boundaries. On the other hand, we took a peak in MCE after the up-sweep and down-sweep crossing points, as shown in Fig.~\ref{fig:PD}.

We identify five distinct magnetic phases alongside the paramagnetic phase. Among these, three phases are aligned with the canonical behavior of a TLAF. These include the 120$^{\circ}$ phase (M1) or Y phase, the UUD phase featuring 1/3 $M_{\text{sat}}$ plateau (M3), and the oblique phase (M5). We first discuss these three phases. However, two additional phases (denoted by M2 and M4) will also be addressed, albeit their precise nature remains unclear at present.

\begin{figure}
	\centering
    \includegraphics[width=1\columnwidth]{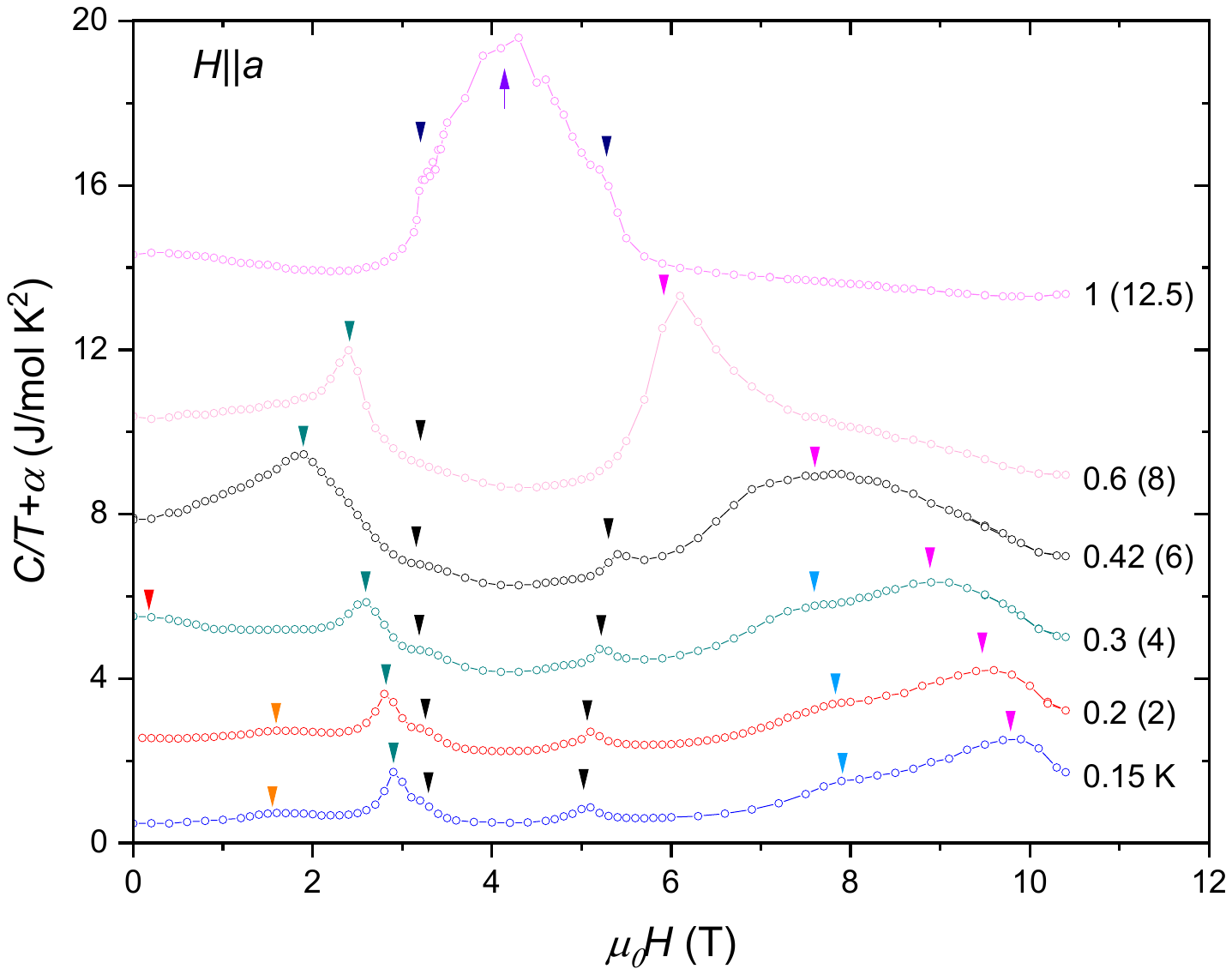}
\caption{\label{fig:HC2}Specific heat divided by temperature, $C/T$ as a function of applied magnetic field, $H$ of single crystal KYbSe$_2$ at different fixed thermal bath temperatures. For clarity, each curve is shifted by a constant, $\alpha$ as indicated in parentheses. Arrows indicate the position of anomalies attributed to transitions between consecutive magnetically ordered states.}
\end{figure}

\begin{figure}
	\centering
    \includegraphics[width=1\columnwidth]{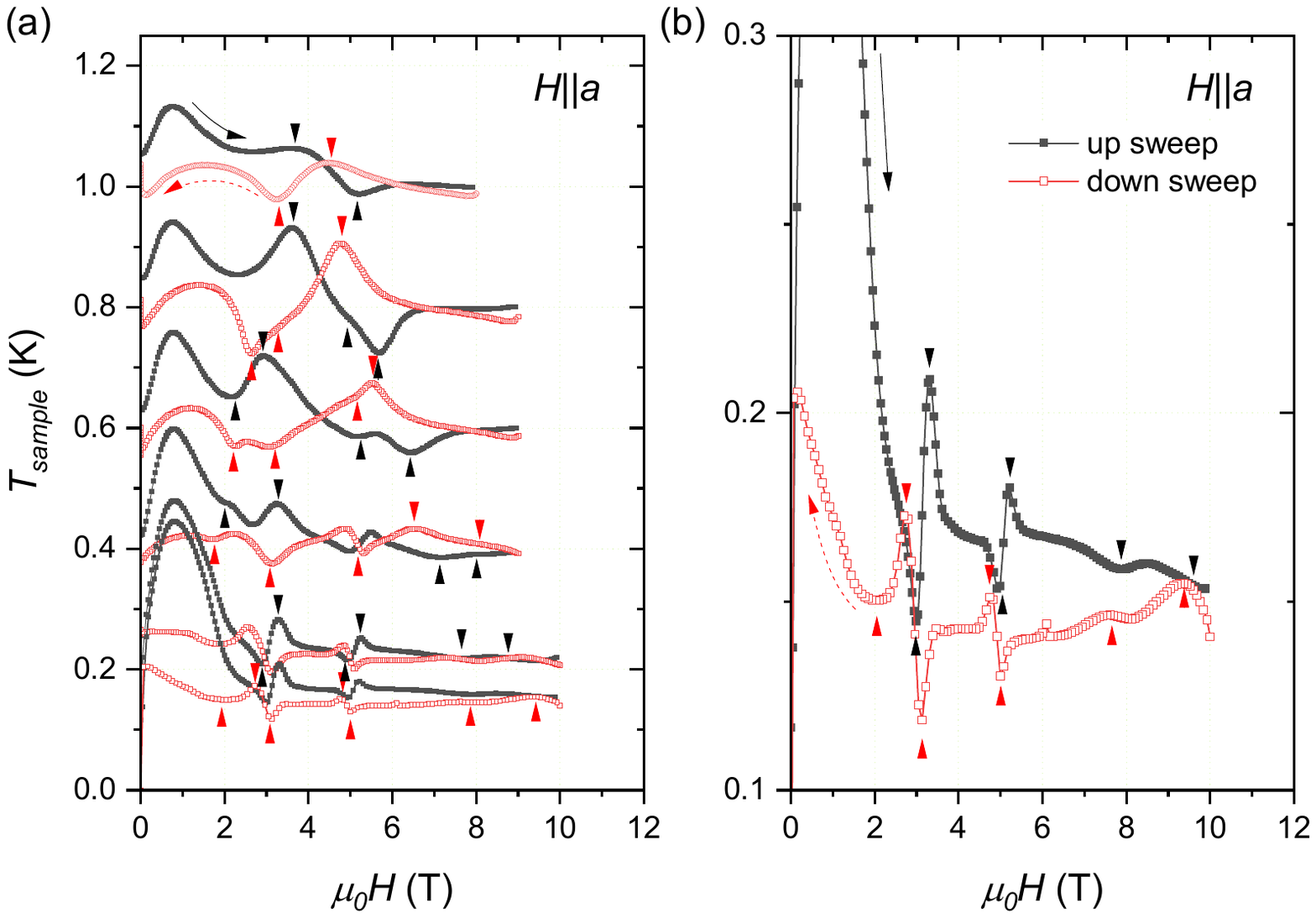}
\caption{\label{fig:mce}Magnetocaloric effect, MCE measurements at various initial temperatures in zero magnetic field. (a) Black solid and red open symbols represent the sample temperature, $T_{sample}$ as a function of up-sweep and down-sweep magnetic field, respectively. (b) A zoomed-in view of the MCE of the sample around the lowest temperature regime. Vertical arrows indicate the anomalies in $T_{sample}$ with sweeping magnetic field.}
\end{figure}

\begin{figure}
	\centering
    \includegraphics[width=1\columnwidth]{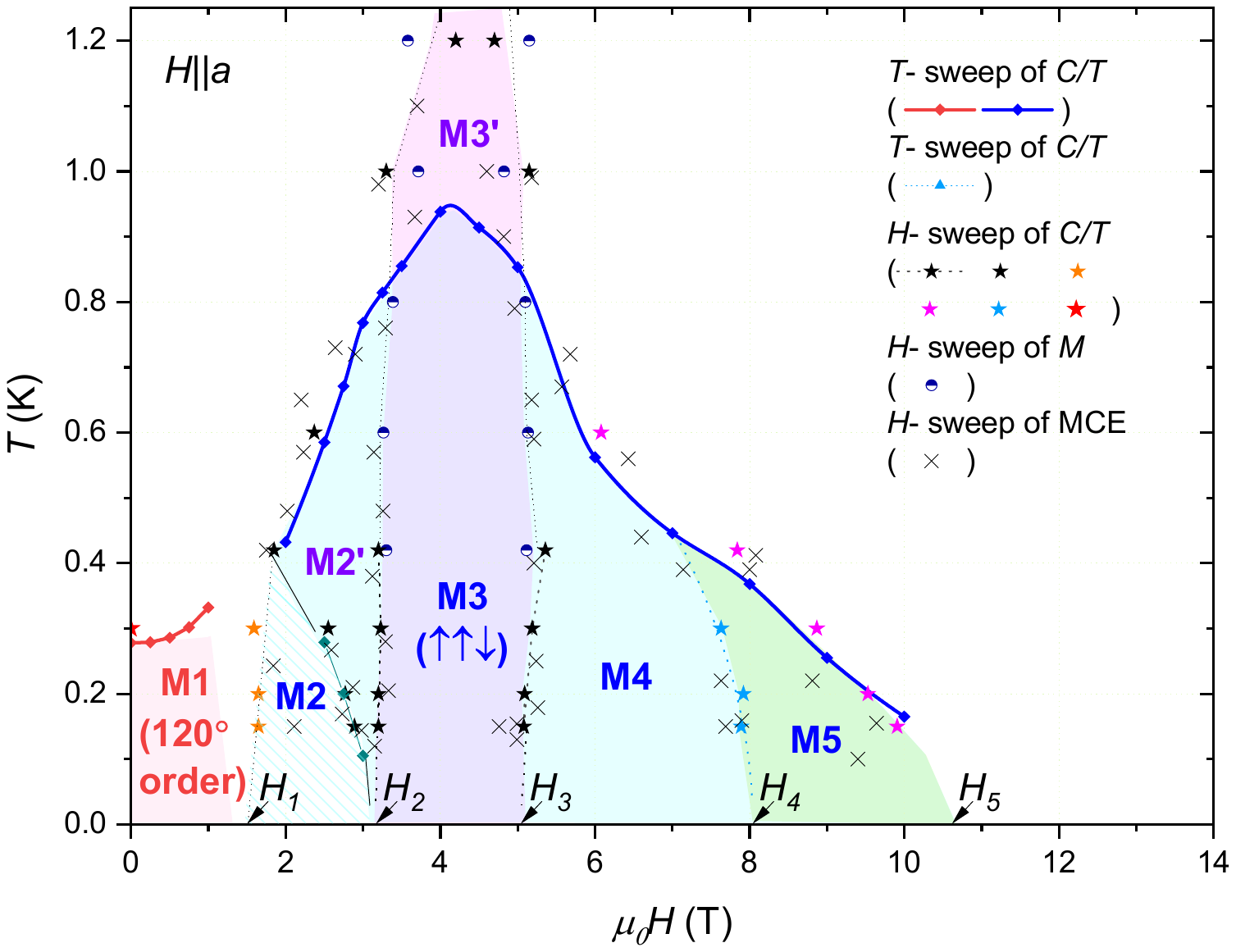}
\caption{(a) The $H-T$ phase diagram of KYbSe$_2$ derived from anomalies in $C/T$ and MCE. The solid lines indicate the boundary of the ordered phase derived from anomalies in the $C/T$($T$). The stars and dotted lines indicate transitions derived from anomalies in $C/T$($H$). The cross symbols indicate anomalies in MCE. The half-filled circles are from data published previously~\cite{xing2019crystal}. We separate the phase diagram into five regions, labeled M1-5, of diﬀerent magnetic orders, with $H_1$-$H_5$ indicating the upper fields of these regions. %(remove it?(b) The residual Sommerfeld coefficient $\gamma_0$ )
\label{fig:PD}}
\end{figure}

\begin{figure}
	\centering
    \includegraphics[width=1\columnwidth]{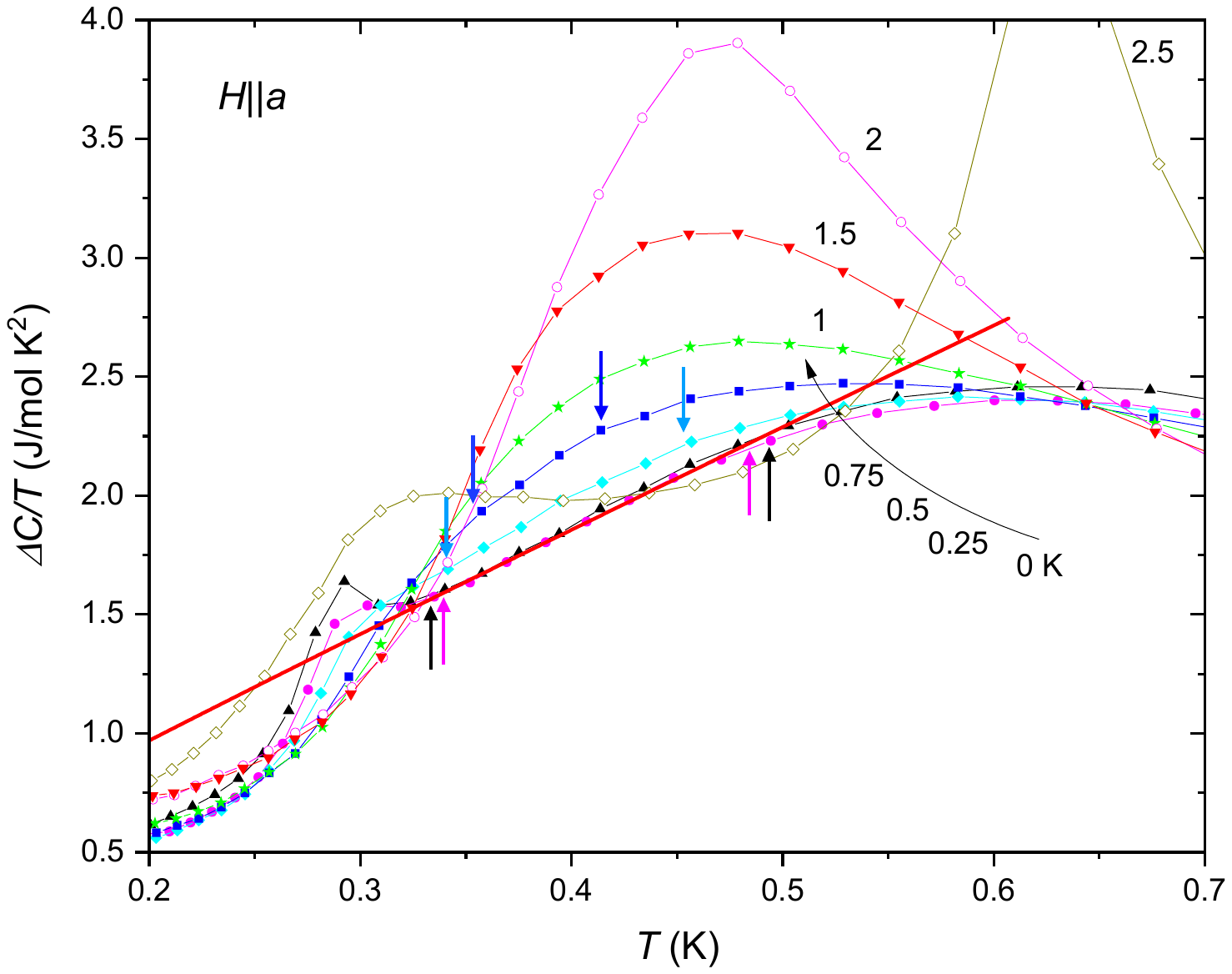}
\caption{A zoomed-in view of $C/T$ around the ordering transition. The solid red line is a guide to the eye for $T$-linear behavior, possibly indicative of a proximate quantum spin liquid state.\label{fig:Linear}}
\end{figure}

We designate M1 as the 120-degree ordered phase and M2 as the Y state. These identifications are based on system-specific calculations~\cite{scheie2023proximate,farnell2009high} and parallels drawn with other triangular lattice antiferromagnets like Ba$_{3}$CoSb$_{2}$O$_{9}$~\cite{susuki2013magnetization,quirion2015magnetic}. In the low-field region, a single magnetic phase transition is observed, implying isotropic magnetic behavior within the plane~\cite{Jolicoeur1990tri}. This suggests that the spin Hamiltonian could be aptly characterized by either XXZ or isotropic Heisenberg interactions. This implies that the long-range ordering at zero magnetic field is primarily propelled by weak interactions between layers, indicating a potential strategy to suppress such long-range magnetic ordering is to engineer the interlayer interactions. Additionally, when analyzing heat capacity at zero field, we determine that the magnetic moment of Yb$^{3+}$ has an upper bound of approximately 0.6~$\mu_{B}$. This value is notably diminished due to the influence of magnetic frustration, distinguishing it from nominal values and those observed in non-frustrated magnetic systems. However, this outcome dovetails with the magnetic moment extracted from other experimental methods. 
In Fig~\ref{fig:Linear}, below 1~T, $C/T$ exhibits a $T$-linear behavior, in the range indicated by vertical arrows. In the fermionic spinon mean-field approach, low-energy dispersion is isotropic and linear at zero field, forming pockets for $H~\neq~0$ with radii proportional to $H$. At the mean-field level, this leads to a transition from $C~\thicksim~T^2$ under zero field to $C~\thicksim~HT$.~\cite{Ran2007linear,Barthelemy2022linear,Liu2021linear,Sindzingre2000linear} 
This aligns with our findings, and Fig.~\ref{fig:HC2} further illustrates the $C/T~\thicksim~H$ behavior for $H <$~1~T at a temperature of 0.42~K.

%Based on the phase transitions we found in C/T(T) and C/T(H), we present the H-T phase diagram of KYbSe2 in Fig. 4. This phase diagram illustrates five consecutive magnetically ordered phases with H||a, labeled as M1-M5. The magnetic fields at which the magnetic phase boundaries are extrapolated to zero temperature are denoted as H1-H5. Additionally, we provide schematics depicting potential magnet orders in each region. 

Previous isothermal magnetization measurements conducted on {\KYS} between $H_2$ ($\sim 3$ T) and $H_3$ ($\sim 5$~T) have revealed a clear plateau at one-third of the saturation magnetization when $H||ab$~\cite{samulon2008ordered}. This plateau corresponds to the phase denoted M3 in our $H-T$ phase diagram. The spin structure of this plateau is expected to be a collinear UUD magnetic order, a prediction supported by various theoretical calculations~\cite{balents2010spin,chen2013ground,syromyatnikov2023unusual,honecker2004magnetization}, and observed in a number of TLAF compounds~\cite{Tsujii2007CCB,Hwang2012uud,lee2014magnetic,Lee2017uud,Gao2022uud,Smirnov2007uud}. This implies that the collinear UUD state is most likely developed in the M3 region in KYbSe$_2$ when $H||a$. One noticeable feature is that the UUD phase maintains its width down to zero temperature. This persistence indicates that quantum fluctuations due to the $S = \frac{1}{2}$ state of Yb$^{3+}$ ions stabilize this plateau phenomenon rather than thermal fluctuations. Meanwhile, the UUD phase of classical spins only exists at a single point at $T = 0$ in the $H-T$ phase diagram, and thermal fluctuations expand the width of the phase with increasing temperature.

In contrast to other TLAF with $S$ = 1/2 with a noticeable single-ion anisotropy, such as Cs$_2$CuBr$_4$~\cite{Tsujii2007CCB}, MCE in Fig.~\ref{fig:mce} for our scenario does not provide clear evidence of a first-order phase transition between the Y phase and the UUD phase. Instead, the transition consistently manifests as a peak and valley structure, characteristic of a second-order phase transition under equilibrium conditions. This observation is further supported by the absence of any magnetization jump when entering or leaving the plateau~\cite{Viv2014mce}. In cases where magnetic anisotropy leads to a spin gap, a first-order phase transition is observed in thermodynamic measurements and magnetization as an abrupt change. Therefore, these observations also support the nearly isotropic magnetism of the title compound.

At higher fields, we observe the emergence of the M4 and M5 phases. However, theoretical investigations \cite{seabra2011phase} have proposed that a single phase with a spin structure known as the oblique phase stabilizes subsequent to the UUD phase before the magnetization saturation. This stabilization is attributed to the subtle interplay between antiferromagnetic interactions and Zeeman energy. Notably, in Fig.~\ref{fig:HC2}, the peaks between the UUD (M3) phase and the M4 phase exhibit distinct sharpness. Conversely, the phase boundaries between M4 and M5 appear weaker. A similar pattern is discernible in MCE data illustrated in Fig.~\ref{fig:mce}. Drawing from these observations, we hypothesize that the M4 phase likely corresponds to the oblique phase, while M5 might represent a phase characterized by a slight change in spin angle relative to the oblique configuration. However, a more in-depth investigation is required to determine the exact nature of the M4 and M5 phases.

The presence of two additional phases, M2' and M5, has also been noted. In canonical TLAFs described by the Heisenberg exchange interaction with nearest neighbor exchange interaction or XXZ Hamiltonian, the three aforementioned phases are typically observed. Thus, the emergence of the M2' and M5 phases suggests the possibility of a novel spin flop phase characterized by spins being tilted away from the theoretically predicted spin structures. The origin of this behavior remains unclear for the time being, but it implies that the additional next-nearest-neighbor magnetic interaction known to exist in the title compound could play a role in stabilizing these extra magnetic phases. Numerous instances exist wherein an additional magnetic interaction term, such as next nearest neighbor exchange interaction ($J_2$), induces the appearance of supplementary phases, as seen in square lattice and cubic lattice systems \cite{Seabra2016square,Pankratova2021cubic}. Moreover, the presence of slight, albeit non-negligible, in-plane anisotropy could potentially trigger a change in spin structures under a magnetic field. Consequently, our study  underscores the need for further investigations to illuminate the role of $J_2$ in stabilizing the new phase, as well as meticulous research to uncover the nuances of the in-plane anisotropy.

\section{\label{sec:conclusion}Conclusion}
To summarize, we conducted specific heat and magnetocaloric effect (MCE) experiments up to 11 T at low temperatures ($T$ $\geq$ 30~mK). We observe the single magnetic phase transition to 120$^{\circ}$ at zero magnetic field, which indicates negligible in-plane magnetic anisotropy. Notably, we observed $T$-linear and linear increasing features in $C/T$ above the long-range magnetic phase transition, which serves as a signature of a proximate QSL. Upon applying a magnetic field, the $T$-linear feature in $C/T$ rapidly suppresses. Furthermore, our investigation unveiled a sequence of 2nd order phase transitions from the 120$^{\circ}$ phase to the UUD phase and subsequently to the oblique phase. These phases are emblematic of isotropic TLAFs. Additionally, we uncovered the existence of two new phases, urging careful consideration in the formulation of a spin Hamiltonian model to fully comprehend this compound's behavior. 
%Moreover, two more ordered phases were identified at finite temperatures. The boundaries of the M3 phase, suggested as having a collinear up-up-down spin configuration with a plateau feature in magnetization and the static magnetic moment of Yb3+ estimated through analysis of the Schottky tail. Furthermore, local maxima of γ0 in the magnetic field suggest that Hi are quantum critical points corresponding to second-order phase transitions.

\section*{Acknowledgements} \label{sec:acknowledgements}
    We acknowledge a fruitful discussion with C. D. Batista. S. L., A. J. W. and R. M. acknowledge LDRD program at Los Alamos National Laboratory. This material is based upon work supported by the US Department of Energy, Office of Science, National Quantum Information Science Research Centers, Quantum Science Center. M. L. and S. Z. were funded by QSC to perform data analysis and manuscript writing. A portion of this work was performed at the National High Magnetic Field Laboratory, which is supported by National Science Foundation Cooperative Agreement No. DMR-2128556, the State of Florida, and the U.S. Department of Energy.

\bibliography{Main_v58}% Produces the bibliography via BibTeX.

\begin{thebibliography}{65}%
\makeatletter
\providecommand \@ifxundefined [1]{%
 \@ifx{#1\undefined}
}%
\providecommand \@ifnum [1]{%
 \ifnum #1\expandafter \@firstoftwo
 \else \expandafter \@secondoftwo
 \fi
}%
\providecommand \@ifx [1]{%
 \ifx #1\expandafter \@firstoftwo
 \else \expandafter \@secondoftwo
 \fi
}%
\providecommand \natexlab [1]{#1}%
\providecommand \enquote  [1]{``#1''}%
\providecommand \bibnamefont  [1]{#1}%
\providecommand \bibfnamefont [1]{#1}%
\providecommand \citenamefont [1]{#1}%
\providecommand \href@noop [0]{\@secondoftwo}%
\providecommand \href [0]{\begingroup \@sanitize@url \@href}%
\providecommand \@href[1]{\@@startlink{#1}\@@href}%
\providecommand \@@href[1]{\endgroup#1\@@endlink}%
\providecommand \@sanitize@url [0]{\catcode `\\12\catcode `\$12\catcode
  `\&12\catcode `\#12\catcode `\^12\catcode `\_12\catcode `\%12\relax}%
\providecommand \@@startlink[1]{}%
\providecommand \@@endlink[0]{}%
\providecommand \url  [0]{\begingroup\@sanitize@url \@url }%
\providecommand \@url [1]{\endgroup\@href {#1}{\urlprefix }}%
\providecommand \urlprefix  [0]{URL }%
\providecommand \Eprint [0]{\href }%
\providecommand \doibase [0]{https://doi.org/}%
\providecommand \selectlanguage [0]{\@gobble}%
\providecommand \bibinfo  [0]{\@secondoftwo}%
\providecommand \bibfield  [0]{\@secondoftwo}%
\providecommand \translation [1]{[#1]}%
\providecommand \BibitemOpen [0]{}%
\providecommand \bibitemStop [0]{}%
\providecommand \bibitemNoStop [0]{.\EOS\space}%
\providecommand \EOS [0]{\spacefactor3000\relax}%
\providecommand \BibitemShut  [1]{\csname bibitem#1\endcsname}%
\let\auto@bib@innerbib\@empty
%</preamble>
\bibitem [{\citenamefont {Balents}(2010)}]{balents2010spin}%
  \BibitemOpen
  \bibfield  {author} {\bibinfo {author} {\bibfnamefont {L.}~\bibnamefont
  {Balents}},\ }\bibfield  {title} {\bibinfo {title} {Spin liquids in
  frustrated magnets},\ }\href@noop {} {\bibfield  {journal} {\bibinfo
  {journal} {nature}\ }\textbf {\bibinfo {volume} {464}},\ \bibinfo {pages}
  {199} (\bibinfo {year} {2010})}\BibitemShut {NoStop}%
\bibitem [{\citenamefont {Savary}\ and\ \citenamefont
  {Balents}(2016)}]{savary2016quantum}%
  \BibitemOpen
  \bibfield  {author} {\bibinfo {author} {\bibfnamefont {L.}~\bibnamefont
  {Savary}}\ and\ \bibinfo {author} {\bibfnamefont {L.}~\bibnamefont
  {Balents}},\ }\bibfield  {title} {\bibinfo {title} {Quantum spin liquids: a
  review},\ }\href@noop {} {\bibfield  {journal} {\bibinfo  {journal} {Reports
  on Progress in Physics}\ }\textbf {\bibinfo {volume} {80}},\ \bibinfo {pages}
  {016502} (\bibinfo {year} {2016})}\BibitemShut {NoStop}%
\bibitem [{\citenamefont {Broholm}\ \emph {et~al.}(2020)\citenamefont
  {Broholm}, \citenamefont {Cava}, \citenamefont {Kivelson}, \citenamefont
  {Nocera}, \citenamefont {Norman},\ and\ \citenamefont
  {Senthil}}]{broholm2020quantum}%
  \BibitemOpen
  \bibfield  {author} {\bibinfo {author} {\bibfnamefont {C.}~\bibnamefont
  {Broholm}}, \bibinfo {author} {\bibfnamefont {R.}~\bibnamefont {Cava}},
  \bibinfo {author} {\bibfnamefont {S.}~\bibnamefont {Kivelson}}, \bibinfo
  {author} {\bibfnamefont {D.}~\bibnamefont {Nocera}}, \bibinfo {author}
  {\bibfnamefont {M.}~\bibnamefont {Norman}},\ and\ \bibinfo {author}
  {\bibfnamefont {T.}~\bibnamefont {Senthil}},\ }\bibfield  {title} {\bibinfo
  {title} {Quantum spin liquids},\ }\href@noop {} {\bibfield  {journal}
  {\bibinfo  {journal} {Science}\ }\textbf {\bibinfo {volume} {367}},\ \bibinfo
  {pages} {eaay0668} (\bibinfo {year} {2020})}\BibitemShut {NoStop}%
\bibitem [{\citenamefont {Zhou}\ \emph {et~al.}(2017)\citenamefont {Zhou},
  \citenamefont {Kanoda},\ and\ \citenamefont {Ng}}]{zhou2017quantum}%
  \BibitemOpen
  \bibfield  {author} {\bibinfo {author} {\bibfnamefont {Y.}~\bibnamefont
  {Zhou}}, \bibinfo {author} {\bibfnamefont {K.}~\bibnamefont {Kanoda}},\ and\
  \bibinfo {author} {\bibfnamefont {T.-K.}\ \bibnamefont {Ng}},\ }\bibfield
  {title} {\bibinfo {title} {Quantum spin liquid states},\ }\href@noop {}
  {\bibfield  {journal} {\bibinfo  {journal} {Reviews of Modern Physics}\
  }\textbf {\bibinfo {volume} {89}},\ \bibinfo {pages} {025003} (\bibinfo
  {year} {2017})}\BibitemShut {NoStop}%
\bibitem [{\citenamefont {Nayak}\ \emph {et~al.}(2008)\citenamefont {Nayak},
  \citenamefont {Simon}, \citenamefont {Stern}, \citenamefont {Freedman},\ and\
  \citenamefont {Sarma}}]{nayak2008non}%
  \BibitemOpen
  \bibfield  {author} {\bibinfo {author} {\bibfnamefont {C.}~\bibnamefont
  {Nayak}}, \bibinfo {author} {\bibfnamefont {S.~H.}\ \bibnamefont {Simon}},
  \bibinfo {author} {\bibfnamefont {A.}~\bibnamefont {Stern}}, \bibinfo
  {author} {\bibfnamefont {M.}~\bibnamefont {Freedman}},\ and\ \bibinfo
  {author} {\bibfnamefont {S.~D.}\ \bibnamefont {Sarma}},\ }\bibfield  {title}
  {\bibinfo {title} {Non-abelian anyons and topological quantum computation},\
  }\href@noop {} {\bibfield  {journal} {\bibinfo  {journal} {Reviews of Modern
  Physics}\ }\textbf {\bibinfo {volume} {80}},\ \bibinfo {pages} {1083}
  (\bibinfo {year} {2008})}\BibitemShut {NoStop}%
\bibitem [{\citenamefont {Kitaev}(2003)}]{kitaev2003fault}%
  \BibitemOpen
  \bibfield  {author} {\bibinfo {author} {\bibfnamefont {A.~Y.}\ \bibnamefont
  {Kitaev}},\ }\bibfield  {title} {\bibinfo {title} {Fault-tolerant quantum
  computation by anyons},\ }\href@noop {} {\bibfield  {journal} {\bibinfo
  {journal} {Annals of physics}\ }\textbf {\bibinfo {volume} {303}},\ \bibinfo
  {pages} {2} (\bibinfo {year} {2003})}\BibitemShut {NoStop}%
\bibitem [{\citenamefont {Anderson}(1973)}]{anderson1973resonating}%
  \BibitemOpen
  \bibfield  {author} {\bibinfo {author} {\bibfnamefont {P.~W.}\ \bibnamefont
  {Anderson}},\ }\bibfield  {title} {\bibinfo {title} {Resonating valence
  bonds: {A} new kind of insulator?},\ }\href@noop {} {\bibfield  {journal}
  {\bibinfo  {journal} {Materials Research Bulletin}\ }\textbf {\bibinfo
  {volume} {8}},\ \bibinfo {pages} {153} (\bibinfo {year} {1973})}\BibitemShut
  {NoStop}%
\bibitem [{\citenamefont {Moessner}\ and\ \citenamefont
  {Ramirez}(2006)}]{moessner2006geometrical}%
  \BibitemOpen
  \bibfield  {author} {\bibinfo {author} {\bibfnamefont {R.}~\bibnamefont
  {Moessner}}\ and\ \bibinfo {author} {\bibfnamefont {A.~P.}\ \bibnamefont
  {Ramirez}},\ }\bibfield  {title} {\bibinfo {title} {Geometrical
  frustration},\ }\href@noop {} {\bibfield  {journal} {\bibinfo  {journal}
  {Physics Today}\ }\textbf {\bibinfo {volume} {59}},\ \bibinfo {pages} {24}
  (\bibinfo {year} {2006})}\BibitemShut {NoStop}%
\bibitem [{\citenamefont {Hu}\ \emph {et~al.}(2019)\citenamefont {Hu},
  \citenamefont {Zhu}, \citenamefont {Eggert},\ and\ \citenamefont
  {He}}]{hu2019dirac}%
  \BibitemOpen
  \bibfield  {author} {\bibinfo {author} {\bibfnamefont {S.}~\bibnamefont
  {Hu}}, \bibinfo {author} {\bibfnamefont {W.}~\bibnamefont {Zhu}}, \bibinfo
  {author} {\bibfnamefont {S.}~\bibnamefont {Eggert}},\ and\ \bibinfo {author}
  {\bibfnamefont {Y.-C.}\ \bibnamefont {He}},\ }\bibfield  {title} {\bibinfo
  {title} {Dirac spin liquid on the spin-1/2 triangular {H}eisenberg
  antiferromagnet},\ }\href@noop {} {\bibfield  {journal} {\bibinfo  {journal}
  {Physical review letters}\ }\textbf {\bibinfo {volume} {123}},\ \bibinfo
  {pages} {207203} (\bibinfo {year} {2019})}\BibitemShut {NoStop}%
\bibitem [{\citenamefont {Gong}\ \emph {et~al.}(2019)\citenamefont {Gong},
  \citenamefont {Zheng}, \citenamefont {Lee}, \citenamefont {Lu},\ and\
  \citenamefont {Sheng}}]{gong2019chiral}%
  \BibitemOpen
  \bibfield  {author} {\bibinfo {author} {\bibfnamefont {S.-S.}\ \bibnamefont
  {Gong}}, \bibinfo {author} {\bibfnamefont {W.}~\bibnamefont {Zheng}},
  \bibinfo {author} {\bibfnamefont {M.}~\bibnamefont {Lee}}, \bibinfo {author}
  {\bibfnamefont {Y.-M.}\ \bibnamefont {Lu}},\ and\ \bibinfo {author}
  {\bibfnamefont {D.}~\bibnamefont {Sheng}},\ }\bibfield  {title} {\bibinfo
  {title} {Chiral spin liquid with spinon {F}ermi surfaces in the spin-1 2
  triangular {H}eisenberg model},\ }\href@noop {} {\bibfield  {journal}
  {\bibinfo  {journal} {Physical Review B}\ }\textbf {\bibinfo {volume}
  {100}},\ \bibinfo {pages} {241111} (\bibinfo {year} {2019})}\BibitemShut
  {NoStop}%
\bibitem [{\citenamefont {Manuel}\ and\ \citenamefont
  {Ceccatto}(1999)}]{manuel1999magnetic}%
  \BibitemOpen
  \bibfield  {author} {\bibinfo {author} {\bibfnamefont {L.}~\bibnamefont
  {Manuel}}\ and\ \bibinfo {author} {\bibfnamefont {H.}~\bibnamefont
  {Ceccatto}},\ }\bibfield  {title} {\bibinfo {title} {Magnetic and quantum
  disordered phases in triangular-lattice {H}eisenberg antiferromagnets},\
  }\href@noop {} {\bibfield  {journal} {\bibinfo  {journal} {Physical Review
  B}\ }\textbf {\bibinfo {volume} {60}},\ \bibinfo {pages} {9489} (\bibinfo
  {year} {1999})}\BibitemShut {NoStop}%
\bibitem [{\citenamefont {Kaneko}\ \emph {et~al.}(2014)\citenamefont {Kaneko},
  \citenamefont {Morita},\ and\ \citenamefont {Imada}}]{kaneko2014gapless}%
  \BibitemOpen
  \bibfield  {author} {\bibinfo {author} {\bibfnamefont {R.}~\bibnamefont
  {Kaneko}}, \bibinfo {author} {\bibfnamefont {S.}~\bibnamefont {Morita}},\
  and\ \bibinfo {author} {\bibfnamefont {M.}~\bibnamefont {Imada}},\ }\bibfield
   {title} {\bibinfo {title} {Gapless spin-liquid phase in an extended spin 1/2
  triangular {H}eisenberg model},\ }\href@noop {} {\bibfield  {journal}
  {\bibinfo  {journal} {Journal of the Physical Society of Japan}\ }\textbf
  {\bibinfo {volume} {83}},\ \bibinfo {pages} {093707} (\bibinfo {year}
  {2014})}\BibitemShut {NoStop}%
\bibitem [{\citenamefont {Iqbal}\ \emph {et~al.}(2016)\citenamefont {Iqbal},
  \citenamefont {Hu}, \citenamefont {Thomale}, \citenamefont {Poilblanc},\ and\
  \citenamefont {Becca}}]{iqbal2016spin}%
  \BibitemOpen
  \bibfield  {author} {\bibinfo {author} {\bibfnamefont {Y.}~\bibnamefont
  {Iqbal}}, \bibinfo {author} {\bibfnamefont {W.-J.}\ \bibnamefont {Hu}},
  \bibinfo {author} {\bibfnamefont {R.}~\bibnamefont {Thomale}}, \bibinfo
  {author} {\bibfnamefont {D.}~\bibnamefont {Poilblanc}},\ and\ \bibinfo
  {author} {\bibfnamefont {F.}~\bibnamefont {Becca}},\ }\bibfield  {title}
  {\bibinfo {title} {Spin liquid nature in the {H}eisenberg {$J_{1}-J_{2}$}
  triangular antiferromagnet},\ }\href@noop {} {\bibfield  {journal} {\bibinfo
  {journal} {Physical Review B}\ }\textbf {\bibinfo {volume} {93}},\ \bibinfo
  {pages} {144411} (\bibinfo {year} {2016})}\BibitemShut {NoStop}%
\bibitem [{\citenamefont {Li}\ \emph {et~al.}(2015)\citenamefont {Li},
  \citenamefont {Bishop},\ and\ \citenamefont
  {Campbell}}]{li2015quasiclassical}%
  \BibitemOpen
  \bibfield  {author} {\bibinfo {author} {\bibfnamefont {P.~H.}\ \bibnamefont
  {Li}}, \bibinfo {author} {\bibfnamefont {R.~F.}\ \bibnamefont {Bishop}},\
  and\ \bibinfo {author} {\bibfnamefont {C.~E.}\ \bibnamefont {Campbell}},\
  }\bibfield  {title} {\bibinfo {title} {Quasiclassical magnetic order and its
  loss in a spin-{$\frac{1}{2}$} {H}eisenberg antiferromagnet on a triangular
  lattice with competing bonds},\ }\href@noop {} {\bibfield  {journal}
  {\bibinfo  {journal} {Physical Review B}\ }\textbf {\bibinfo {volume} {91}},\
  \bibinfo {pages} {014426} (\bibinfo {year} {2015})}\BibitemShut {NoStop}%
\bibitem [{\citenamefont {Hu}\ \emph {et~al.}(2015)\citenamefont {Hu},
  \citenamefont {Gong}, \citenamefont {Zhu},\ and\ \citenamefont
  {Sheng}}]{hu2015competing}%
  \BibitemOpen
  \bibfield  {author} {\bibinfo {author} {\bibfnamefont {W.-J.}\ \bibnamefont
  {Hu}}, \bibinfo {author} {\bibfnamefont {S.-S.}\ \bibnamefont {Gong}},
  \bibinfo {author} {\bibfnamefont {W.}~\bibnamefont {Zhu}},\ and\ \bibinfo
  {author} {\bibfnamefont {D.}~\bibnamefont {Sheng}},\ }\bibfield  {title}
  {\bibinfo {title} {Competing spin-liquid states in the spin-{$\frac{1}{2}$}
  {H}eisenberg model on the triangular lattice},\ }\href@noop {} {\bibfield
  {journal} {\bibinfo  {journal} {Physical Review B}\ }\textbf {\bibinfo
  {volume} {92}},\ \bibinfo {pages} {140403} (\bibinfo {year}
  {2015})}\BibitemShut {NoStop}%
\bibitem [{\citenamefont {Saadatmand}\ and\ \citenamefont
  {McCulloch}(2016)}]{saadatmand2016symmetry}%
  \BibitemOpen
  \bibfield  {author} {\bibinfo {author} {\bibfnamefont {S.}~\bibnamefont
  {Saadatmand}}\ and\ \bibinfo {author} {\bibfnamefont {I.}~\bibnamefont
  {McCulloch}},\ }\bibfield  {title} {\bibinfo {title} {Symmetry
  fractionalization in the topological phase of the spin-{$\frac{1}{2}$}
  {$J_{1}- J_{2}$} triangular {H}eisenberg model},\ }\href@noop {} {\bibfield
  {journal} {\bibinfo  {journal} {Physical Review B}\ }\textbf {\bibinfo
  {volume} {94}},\ \bibinfo {pages} {121111} (\bibinfo {year}
  {2016})}\BibitemShut {NoStop}%
\bibitem [{\citenamefont {Wietek}\ and\ \citenamefont
  {L{\"a}uchli}(2017)}]{wietek2017chiral}%
  \BibitemOpen
  \bibfield  {author} {\bibinfo {author} {\bibfnamefont {A.}~\bibnamefont
  {Wietek}}\ and\ \bibinfo {author} {\bibfnamefont {A.~M.}\ \bibnamefont
  {L{\"a}uchli}},\ }\bibfield  {title} {\bibinfo {title} {Chiral spin liquid
  and quantum criticality in extended {$S = \frac{1}{2}$} {H}eisenberg models
  on the triangular lattice},\ }\href@noop {} {\bibfield  {journal} {\bibinfo
  {journal} {Physical Review B}\ }\textbf {\bibinfo {volume} {95}},\ \bibinfo
  {pages} {035141} (\bibinfo {year} {2017})}\BibitemShut {NoStop}%
\bibitem [{\citenamefont {Gong}\ \emph {et~al.}(2017)\citenamefont {Gong},
  \citenamefont {Zhu}, \citenamefont {Zhu}, \citenamefont {Sheng},\ and\
  \citenamefont {Yang}}]{gong2017global}%
  \BibitemOpen
  \bibfield  {author} {\bibinfo {author} {\bibfnamefont {S.-S.}\ \bibnamefont
  {Gong}}, \bibinfo {author} {\bibfnamefont {W.}~\bibnamefont {Zhu}}, \bibinfo
  {author} {\bibfnamefont {J.-X.}\ \bibnamefont {Zhu}}, \bibinfo {author}
  {\bibfnamefont {D.~N.}\ \bibnamefont {Sheng}},\ and\ \bibinfo {author}
  {\bibfnamefont {K.}~\bibnamefont {Yang}},\ }\bibfield  {title} {\bibinfo
  {title} {Global phase diagram and quantum spin liquids in a
  spin-{$\frac{1}{2}$} triangular antiferromagnet},\ }\href@noop {} {\bibfield
  {journal} {\bibinfo  {journal} {Physical Review B}\ }\textbf {\bibinfo
  {volume} {96}},\ \bibinfo {pages} {075116} (\bibinfo {year}
  {2017})}\BibitemShut {NoStop}%
\bibitem [{\citenamefont {Bauer}\ and\ \citenamefont
  {Fj{\ae}restad}(2017)}]{bauer2017schwinger}%
  \BibitemOpen
  \bibfield  {author} {\bibinfo {author} {\bibfnamefont {D.-V.}\ \bibnamefont
  {Bauer}}\ and\ \bibinfo {author} {\bibfnamefont {J.}~\bibnamefont
  {Fj{\ae}restad}},\ }\bibfield  {title} {\bibinfo {title} {Schwinger-boson
  mean-field study of the {$J_{1}- J_{2}$} {H}eisenberg quantum antiferromagnet
  on the triangular lattice},\ }\href@noop {} {\bibfield  {journal} {\bibinfo
  {journal} {Physical Review B}\ }\textbf {\bibinfo {volume} {96}},\ \bibinfo
  {pages} {165141} (\bibinfo {year} {2017})}\BibitemShut {NoStop}%
\bibitem [{\citenamefont {Nishimori}\ and\ \citenamefont
  {Miyashita}(1986)}]{nishimori1986magnetization}%
  \BibitemOpen
  \bibfield  {author} {\bibinfo {author} {\bibfnamefont {H.}~\bibnamefont
  {Nishimori}}\ and\ \bibinfo {author} {\bibfnamefont {S.}~\bibnamefont
  {Miyashita}},\ }\bibfield  {title} {\bibinfo {title} {Magnetization process
  of the spin-1/2 antiferromagnetic {I}sing-like {H}eisenberg model on the
  triangular lattice},\ }\href@noop {} {\bibfield  {journal} {\bibinfo
  {journal} {Journal of the Physical Society of Japan}\ }\textbf {\bibinfo
  {volume} {55}},\ \bibinfo {pages} {4448} (\bibinfo {year}
  {1986})}\BibitemShut {NoStop}%
\bibitem [{\citenamefont {Chubukov}\ and\ \citenamefont
  {Golosov}(1991)}]{chubukov1991quantum}%
  \BibitemOpen
  \bibfield  {author} {\bibinfo {author} {\bibfnamefont {A.}~\bibnamefont
  {Chubukov}}\ and\ \bibinfo {author} {\bibfnamefont {D.}~\bibnamefont
  {Golosov}},\ }\bibfield  {title} {\bibinfo {title} {Quantum theory of an
  antiferromagnet on a triangular lattice in a magnetic field},\ }\href@noop {}
  {\bibfield  {journal} {\bibinfo  {journal} {Journal of Physics: Condensed
  Matter}\ }\textbf {\bibinfo {volume} {3}},\ \bibinfo {pages} {69} (\bibinfo
  {year} {1991})}\BibitemShut {NoStop}%
\bibitem [{\citenamefont {Susuki}\ \emph {et~al.}(2013)\citenamefont {Susuki},
  \citenamefont {Kurita}, \citenamefont {Tanaka}, \citenamefont {Nojiri},
  \citenamefont {Matsuo}, \citenamefont {Kindo},\ and\ \citenamefont
  {Tanaka}}]{susuki2013magnetization}%
  \BibitemOpen
  \bibfield  {author} {\bibinfo {author} {\bibfnamefont {T.}~\bibnamefont
  {Susuki}}, \bibinfo {author} {\bibfnamefont {N.}~\bibnamefont {Kurita}},
  \bibinfo {author} {\bibfnamefont {T.}~\bibnamefont {Tanaka}}, \bibinfo
  {author} {\bibfnamefont {H.}~\bibnamefont {Nojiri}}, \bibinfo {author}
  {\bibfnamefont {A.}~\bibnamefont {Matsuo}}, \bibinfo {author} {\bibfnamefont
  {K.}~\bibnamefont {Kindo}},\ and\ \bibinfo {author} {\bibfnamefont
  {H.}~\bibnamefont {Tanaka}},\ }\bibfield  {title} {\bibinfo {title}
  {Magnetization process and collective excitations in the {S}= 1/2
  triangular-lattice {H}eisenberg antiferromagnet {Ba$_3$CoSb$_2$O$_9$}},\
  }\href@noop {} {\bibfield  {journal} {\bibinfo  {journal} {Physical Review
  Letters}\ }\textbf {\bibinfo {volume} {110}},\ \bibinfo {pages} {267201}
  (\bibinfo {year} {2013})}\BibitemShut {NoStop}%
\bibitem [{\citenamefont {Villain}\ \emph {et~al.}(1980)\citenamefont
  {Villain}, \citenamefont {Bidaux}, \citenamefont {Carton},\ and\
  \citenamefont {Conte}}]{villain1980order}%
  \BibitemOpen
  \bibfield  {author} {\bibinfo {author} {\bibfnamefont {J.}~\bibnamefont
  {Villain}}, \bibinfo {author} {\bibfnamefont {R.}~\bibnamefont {Bidaux}},
  \bibinfo {author} {\bibfnamefont {J.-P.}\ \bibnamefont {Carton}},\ and\
  \bibinfo {author} {\bibfnamefont {R.}~\bibnamefont {Conte}},\ }\bibfield
  {title} {\bibinfo {title} {Order as an effect of disorder},\ }\href@noop {}
  {\bibfield  {journal} {\bibinfo  {journal} {Journal de Physique}\ }\textbf
  {\bibinfo {volume} {41}},\ \bibinfo {pages} {1263} (\bibinfo {year}
  {1980})}\BibitemShut {NoStop}%
\bibitem [{\citenamefont {Seabra}\ \emph {et~al.}(2011)\citenamefont {Seabra},
  \citenamefont {Momoi}, \citenamefont {Sindzingre},\ and\ \citenamefont
  {Shannon}}]{seabra2011phase}%
  \BibitemOpen
  \bibfield  {author} {\bibinfo {author} {\bibfnamefont {L.}~\bibnamefont
  {Seabra}}, \bibinfo {author} {\bibfnamefont {T.}~\bibnamefont {Momoi}},
  \bibinfo {author} {\bibfnamefont {P.}~\bibnamefont {Sindzingre}},\ and\
  \bibinfo {author} {\bibfnamefont {N.}~\bibnamefont {Shannon}},\ }\bibfield
  {title} {\bibinfo {title} {Phase diagram of the classical {H}eisenberg
  antiferromagnet on a triangular lattice in an applied magnetic field},\
  }\href@noop {} {\bibfield  {journal} {\bibinfo  {journal} {Physical Review
  B}\ }\textbf {\bibinfo {volume} {84}},\ \bibinfo {pages} {214418} (\bibinfo
  {year} {2011})}\BibitemShut {NoStop}%
\bibitem [{\citenamefont {Lee}\ \emph {et~al.}(2014{\natexlab{a}})\citenamefont
  {Lee}, \citenamefont {Choi}, \citenamefont {Huang}, \citenamefont {Ma},
  \citenamefont {Cruz}, \citenamefont {Matsuda}, \citenamefont {Tian},
  \citenamefont {Dun}, \citenamefont {Dong},\ and\ \citenamefont
  {Zhou}}]{lee2014magnetic}%
  \BibitemOpen
  \bibfield  {author} {\bibinfo {author} {\bibfnamefont {M.}~\bibnamefont
  {Lee}}, \bibinfo {author} {\bibfnamefont {E.}~\bibnamefont {Choi}}, \bibinfo
  {author} {\bibfnamefont {X.}~\bibnamefont {Huang}}, \bibinfo {author}
  {\bibfnamefont {J.}~\bibnamefont {Ma}}, \bibinfo {author} {\bibfnamefont
  {C.~D.}\ \bibnamefont {Cruz}}, \bibinfo {author} {\bibfnamefont
  {M.}~\bibnamefont {Matsuda}}, \bibinfo {author} {\bibfnamefont
  {W.}~\bibnamefont {Tian}}, \bibinfo {author} {\bibfnamefont {Z.}~\bibnamefont
  {Dun}}, \bibinfo {author} {\bibfnamefont {S.}~\bibnamefont {Dong}},\ and\
  \bibinfo {author} {\bibfnamefont {H.}~\bibnamefont {Zhou}},\ }\bibfield
  {title} {\bibinfo {title} {Magnetic phase diagram and multiferroicity of
  {Ba$_3$MnNb$_2$O$_9$}: {A} spin-5/2 triangular lattice antiferromagnet with
  weak easy-axis anisotropy},\ }\href@noop {} {\bibfield  {journal} {\bibinfo
  {journal} {Physical Review B}\ }\textbf {\bibinfo {volume} {90}},\ \bibinfo
  {pages} {224402} (\bibinfo {year} {2014}{\natexlab{a}})}\BibitemShut
  {NoStop}%
\bibitem [{\citenamefont {Quirion}\ \emph {et~al.}(2015)\citenamefont
  {Quirion}, \citenamefont {Lapointe-Major}, \citenamefont {Poirier},
  \citenamefont {Quilliam}, \citenamefont {Dun},\ and\ \citenamefont
  {Zhou}}]{quirion2015magnetic}%
  \BibitemOpen
  \bibfield  {author} {\bibinfo {author} {\bibfnamefont {G.}~\bibnamefont
  {Quirion}}, \bibinfo {author} {\bibfnamefont {M.}~\bibnamefont
  {Lapointe-Major}}, \bibinfo {author} {\bibfnamefont {M.}~\bibnamefont
  {Poirier}}, \bibinfo {author} {\bibfnamefont {J.}~\bibnamefont {Quilliam}},
  \bibinfo {author} {\bibfnamefont {Z.}~\bibnamefont {Dun}},\ and\ \bibinfo
  {author} {\bibfnamefont {H.}~\bibnamefont {Zhou}},\ }\bibfield  {title}
  {\bibinfo {title} {Magnetic phase diagram of {Ba$_3$CoSb$_2$O$_9$} as
  determined by ultrasound velocity measurements},\ }\href@noop {} {\bibfield
  {journal} {\bibinfo  {journal} {Physical Review B}\ }\textbf {\bibinfo
  {volume} {92}},\ \bibinfo {pages} {014414} (\bibinfo {year}
  {2015})}\BibitemShut {NoStop}%
\bibitem [{\citenamefont {Lee}\ \emph {et~al.}(2014{\natexlab{b}})\citenamefont
  {Lee}, \citenamefont {Hwang}, \citenamefont {Choi}, \citenamefont {Ma},
  \citenamefont {Cruz}, \citenamefont {Zhu}, \citenamefont {Ke}, \citenamefont
  {Dun},\ and\ \citenamefont {Zhou}}]{lee2014series}%
  \BibitemOpen
  \bibfield  {author} {\bibinfo {author} {\bibfnamefont {M.}~\bibnamefont
  {Lee}}, \bibinfo {author} {\bibfnamefont {J.}~\bibnamefont {Hwang}}, \bibinfo
  {author} {\bibfnamefont {E.}~\bibnamefont {Choi}}, \bibinfo {author}
  {\bibfnamefont {J.}~\bibnamefont {Ma}}, \bibinfo {author} {\bibfnamefont
  {C.~D.}\ \bibnamefont {Cruz}}, \bibinfo {author} {\bibfnamefont
  {M.}~\bibnamefont {Zhu}}, \bibinfo {author} {\bibfnamefont {X.}~\bibnamefont
  {Ke}}, \bibinfo {author} {\bibfnamefont {Z.}~\bibnamefont {Dun}},\ and\
  \bibinfo {author} {\bibfnamefont {H.}~\bibnamefont {Zhou}},\ }\bibfield
  {title} {\bibinfo {title} {Series of phase transitions and multiferroicity in
  the quasi-two-dimensional spin-${\frac{1}{2}}$ triangular-lattice
  antiferromagnet {Ba$_3$CoNb$_2$O$_9$}},\ }\href@noop {} {\bibfield  {journal}
  {\bibinfo  {journal} {Physical Review B}\ }\textbf {\bibinfo {volume} {89}},\
  \bibinfo {pages} {104420} (\bibinfo {year} {2014}{\natexlab{b}})}\BibitemShut
  {NoStop}%
\bibitem [{\citenamefont {Zhou}\ \emph {et~al.}(2012)\citenamefont {Zhou},
  \citenamefont {Xu}, \citenamefont {Hallas}, \citenamefont {Silverstein},
  \citenamefont {Wiebe}, \citenamefont {Umegaki}, \citenamefont {Yan},
  \citenamefont {Murphy}, \citenamefont {Park}, \citenamefont {Qiu} \emph
  {et~al.}}]{zhou2012successive}%
  \BibitemOpen
  \bibfield  {author} {\bibinfo {author} {\bibfnamefont {H.}~\bibnamefont
  {Zhou}}, \bibinfo {author} {\bibfnamefont {C.}~\bibnamefont {Xu}}, \bibinfo
  {author} {\bibfnamefont {A.}~\bibnamefont {Hallas}}, \bibinfo {author}
  {\bibfnamefont {H.}~\bibnamefont {Silverstein}}, \bibinfo {author}
  {\bibfnamefont {C.}~\bibnamefont {Wiebe}}, \bibinfo {author} {\bibfnamefont
  {I.}~\bibnamefont {Umegaki}}, \bibinfo {author} {\bibfnamefont
  {J.}~\bibnamefont {Yan}}, \bibinfo {author} {\bibfnamefont {T.}~\bibnamefont
  {Murphy}}, \bibinfo {author} {\bibfnamefont {J.-H.}\ \bibnamefont {Park}},
  \bibinfo {author} {\bibfnamefont {Y.}~\bibnamefont {Qiu}}, \emph {et~al.},\
  }\bibfield  {title} {\bibinfo {title} {Successive phase transitions and
  extended spin-excitation continuum in the {$S$}= $\frac{1}{2}$
  triangular-lattice antiferromagnet {Ba$_3$CoSb$_2$O$_9$}},\ }\href@noop {}
  {\bibfield  {journal} {\bibinfo  {journal} {Physical Review Letters}\
  }\textbf {\bibinfo {volume} {109}},\ \bibinfo {pages} {267206} (\bibinfo
  {year} {2012})}\BibitemShut {NoStop}%
\bibitem [{\citenamefont {Schmidt}\ \emph {et~al.}(2021)\citenamefont
  {Schmidt}, \citenamefont {Sichelschmidt}, \citenamefont {Ranjith},
  \citenamefont {Doert},\ and\ \citenamefont {Baenitz}}]{schmidt2021yb}%
  \BibitemOpen
  \bibfield  {author} {\bibinfo {author} {\bibfnamefont {B.}~\bibnamefont
  {Schmidt}}, \bibinfo {author} {\bibfnamefont {J.}~\bibnamefont
  {Sichelschmidt}}, \bibinfo {author} {\bibfnamefont {K.}~\bibnamefont
  {Ranjith}}, \bibinfo {author} {\bibfnamefont {T.}~\bibnamefont {Doert}},\
  and\ \bibinfo {author} {\bibfnamefont {M.}~\bibnamefont {Baenitz}},\
  }\bibfield  {title} {\bibinfo {title} {Yb delafossites: {U}nique exchange
  frustration of 4$f$ spin-$\frac{1}{2}$ moments on a perfect triangular
  lattice},\ }\href@noop {} {\bibfield  {journal} {\bibinfo  {journal}
  {Physical Review B}\ }\textbf {\bibinfo {volume} {103}},\ \bibinfo {pages}
  {214445} (\bibinfo {year} {2021})}\BibitemShut {NoStop}%
\bibitem [{\citenamefont {Xing}\ \emph {et~al.}(2019)\citenamefont {Xing},
  \citenamefont {Sanjeewa}, \citenamefont {Kim}, \citenamefont {Stewart},
  \citenamefont {Du}, \citenamefont {Reboredo}, \citenamefont {Custelcean},\
  and\ \citenamefont {Sefat}}]{xing2019crystal}%
  \BibitemOpen
  \bibfield  {author} {\bibinfo {author} {\bibfnamefont {J.}~\bibnamefont
  {Xing}}, \bibinfo {author} {\bibfnamefont {L.~D.}\ \bibnamefont {Sanjeewa}},
  \bibinfo {author} {\bibfnamefont {J.}~\bibnamefont {Kim}}, \bibinfo {author}
  {\bibfnamefont {G.}~\bibnamefont {Stewart}}, \bibinfo {author} {\bibfnamefont
  {M.-H.}\ \bibnamefont {Du}}, \bibinfo {author} {\bibfnamefont {F.~A.}\
  \bibnamefont {Reboredo}}, \bibinfo {author} {\bibfnamefont {R.}~\bibnamefont
  {Custelcean}},\ and\ \bibinfo {author} {\bibfnamefont {A.~S.}\ \bibnamefont
  {Sefat}},\ }\bibfield  {title} {\bibinfo {title} {Crystal synthesis and
  frustrated magnetism in triangular lattice {CsRESe$_2$} {(RE= La--Lu)}:
  Quantum spin liquid candidates {CsCeSe$_2$} and {CsYbSe$_2$}},\ }\href@noop
  {} {\bibfield  {journal} {\bibinfo  {journal} {ACS Materials Letters}\
  }\textbf {\bibinfo {volume} {2}},\ \bibinfo {pages} {71} (\bibinfo {year}
  {2019})}\BibitemShut {NoStop}%
\bibitem [{\citenamefont {Zangeneh}\ \emph {et~al.}(2019)\citenamefont
  {Zangeneh}, \citenamefont {Avdoshenko}, \citenamefont {van~den Brink},\ and\
  \citenamefont {Hozoi}}]{zangeneh2019single}%
  \BibitemOpen
  \bibfield  {author} {\bibinfo {author} {\bibfnamefont {Z.}~\bibnamefont
  {Zangeneh}}, \bibinfo {author} {\bibfnamefont {S.}~\bibnamefont
  {Avdoshenko}}, \bibinfo {author} {\bibfnamefont {J.}~\bibnamefont {van~den
  Brink}},\ and\ \bibinfo {author} {\bibfnamefont {L.}~\bibnamefont {Hozoi}},\
  }\bibfield  {title} {\bibinfo {title} {Single-site magnetic anisotropy
  governed by interlayer cation charge imbalance in triangular-lattice
  {$A$Yb$X$}$_{2}$},\ }\href@noop {} {\bibfield  {journal} {\bibinfo  {journal}
  {Physical Review B}\ }\textbf {\bibinfo {volume} {100}},\ \bibinfo {pages}
  {174436} (\bibinfo {year} {2019})}\BibitemShut {NoStop}%
\bibitem [{\citenamefont {Scheie}\ \emph {et~al.}(2020)\citenamefont {Scheie},
  \citenamefont {Garlea}, \citenamefont {Sanjeewa}, \citenamefont {Xing},\ and\
  \citenamefont {Sefat}}]{scheie2020crystal}%
  \BibitemOpen
  \bibfield  {author} {\bibinfo {author} {\bibfnamefont {A.}~\bibnamefont
  {Scheie}}, \bibinfo {author} {\bibfnamefont {V.~O.}\ \bibnamefont {Garlea}},
  \bibinfo {author} {\bibfnamefont {L.~D.}\ \bibnamefont {Sanjeewa}}, \bibinfo
  {author} {\bibfnamefont {J.}~\bibnamefont {Xing}},\ and\ \bibinfo {author}
  {\bibfnamefont {A.~S.}\ \bibnamefont {Sefat}},\ }\bibfield  {title} {\bibinfo
  {title} {Crystal-field {H}amiltonian and anisotropy in {KErSe$_2$} and
  {CsErSe$_2$}},\ }\href@noop {} {\bibfield  {journal} {\bibinfo  {journal}
  {Physical Review B}\ }\textbf {\bibinfo {volume} {101}},\ \bibinfo {pages}
  {144432} (\bibinfo {year} {2020})}\BibitemShut {NoStop}%
\bibitem [{\citenamefont {Bordelon}\ \emph {et~al.}(2021)\citenamefont
  {Bordelon}, \citenamefont {Liu}, \citenamefont {Posthuma}, \citenamefont
  {Kenney}, \citenamefont {Graf}, \citenamefont {Butch}, \citenamefont
  {Banerjee}, \citenamefont {Calder}, \citenamefont {Balents},\ and\
  \citenamefont {Wilson}}]{bordelon2021frustrated}%
  \BibitemOpen
  \bibfield  {author} {\bibinfo {author} {\bibfnamefont {M.~M.}\ \bibnamefont
  {Bordelon}}, \bibinfo {author} {\bibfnamefont {C.}~\bibnamefont {Liu}},
  \bibinfo {author} {\bibfnamefont {L.}~\bibnamefont {Posthuma}}, \bibinfo
  {author} {\bibfnamefont {E.}~\bibnamefont {Kenney}}, \bibinfo {author}
  {\bibfnamefont {M.}~\bibnamefont {Graf}}, \bibinfo {author} {\bibfnamefont
  {N.}~\bibnamefont {Butch}}, \bibinfo {author} {\bibfnamefont
  {A.}~\bibnamefont {Banerjee}}, \bibinfo {author} {\bibfnamefont
  {S.}~\bibnamefont {Calder}}, \bibinfo {author} {\bibfnamefont
  {L.}~\bibnamefont {Balents}},\ and\ \bibinfo {author} {\bibfnamefont {S.~D.}\
  \bibnamefont {Wilson}},\ }\bibfield  {title} {\bibinfo {title} {Frustrated
  heisenberg {$J_1-J_2$} model within the stretched diamond lattice of
  {LiYbO$_2$}},\ }\href@noop {} {\bibfield  {journal} {\bibinfo  {journal}
  {Physical Review B}\ }\textbf {\bibinfo {volume} {103}},\ \bibinfo {pages}
  {014420} (\bibinfo {year} {2021})}\BibitemShut {NoStop}%
\bibitem [{\citenamefont {Salke}\ \emph {et~al.}(2014)\citenamefont {Salke},
  \citenamefont {Garg}, \citenamefont {Rao}, \citenamefont {Achary},
  \citenamefont {Gupta}, \citenamefont {Mittal},\ and\ \citenamefont
  {Tyagi}}]{salke2014phase}%
  \BibitemOpen
  \bibfield  {author} {\bibinfo {author} {\bibfnamefont {N.~P.}\ \bibnamefont
  {Salke}}, \bibinfo {author} {\bibfnamefont {A.~B.}\ \bibnamefont {Garg}},
  \bibinfo {author} {\bibfnamefont {R.}~\bibnamefont {Rao}}, \bibinfo {author}
  {\bibfnamefont {S.}~\bibnamefont {Achary}}, \bibinfo {author} {\bibfnamefont
  {M.}~\bibnamefont {Gupta}}, \bibinfo {author} {\bibfnamefont
  {R.}~\bibnamefont {Mittal}},\ and\ \bibinfo {author} {\bibfnamefont
  {A.}~\bibnamefont {Tyagi}},\ }\bibfield  {title} {\bibinfo {title} {Phase
  transitions in delafossite {CuLaO$_{2}$} at high pressures},\ }\href@noop {}
  {\bibfield  {journal} {\bibinfo  {journal} {Journal of Applied Physics}\
  }\textbf {\bibinfo {volume} {115}} (\bibinfo {year} {2014})}\BibitemShut
  {NoStop}%
\bibitem [{\citenamefont {Boyraz}\ \emph {et~al.}(2022)\citenamefont {Boyraz},
  \citenamefont {Aksu}, \citenamefont {Guler},\ and\ \citenamefont
  {Arda}}]{boyraz2022effect}%
  \BibitemOpen
  \bibfield  {author} {\bibinfo {author} {\bibfnamefont {C.}~\bibnamefont
  {Boyraz}}, \bibinfo {author} {\bibfnamefont {P.}~\bibnamefont {Aksu}},
  \bibinfo {author} {\bibfnamefont {A.}~\bibnamefont {Guler}},\ and\ \bibinfo
  {author} {\bibfnamefont {L.}~\bibnamefont {Arda}},\ }\bibfield  {title}
  {\bibinfo {title} {The effect of defects formed under pressure on {CuCrO$_2$}
  delafossite},\ }\href@noop {} {\bibfield  {journal} {\bibinfo  {journal} {SN
  Applied Sciences}\ }\textbf {\bibinfo {volume} {4}},\ \bibinfo {pages} {193}
  (\bibinfo {year} {2022})}\BibitemShut {NoStop}%
\bibitem [{\citenamefont {Sichelschmidt}\ \emph {et~al.}(2020)\citenamefont
  {Sichelschmidt}, \citenamefont {Schmidt}, \citenamefont {Schlender},
  \citenamefont {Khim}, \citenamefont {Doert},\ and\ \citenamefont
  {Baenitz}}]{sichelschmidt2020effective}%
  \BibitemOpen
  \bibfield  {author} {\bibinfo {author} {\bibfnamefont {J.}~\bibnamefont
  {Sichelschmidt}}, \bibinfo {author} {\bibfnamefont {B.}~\bibnamefont
  {Schmidt}}, \bibinfo {author} {\bibfnamefont {P.}~\bibnamefont {Schlender}},
  \bibinfo {author} {\bibfnamefont {S.}~\bibnamefont {Khim}}, \bibinfo {author}
  {\bibfnamefont {T.}~\bibnamefont {Doert}},\ and\ \bibinfo {author}
  {\bibfnamefont {M.}~\bibnamefont {Baenitz}},\ }\bibfield  {title} {\bibinfo
  {title} {Effective spin-1/2 moments on a {Yb$^{3+}$} triangular lattice: An
  {ESR} study},\ }\href@noop {} {\bibfield  {journal} {\bibinfo  {journal} {JPS
  Conference Proceedings}\ }\textbf {\bibinfo {volume} {30}},\ \bibinfo {pages}
  {011096} (\bibinfo {year} {2020})}\BibitemShut {NoStop}%
\bibitem [{\citenamefont {Wu}\ \emph {et~al.}(2019)\citenamefont {Wu},
  \citenamefont {Nikitin}, \citenamefont {Wang}, \citenamefont {Zhu},
  \citenamefont {Batista}, \citenamefont {Tsvelik}, \citenamefont {Samarakoon},
  \citenamefont {Tennant}, \citenamefont {Brando}, \citenamefont {Vasylechko}
  \emph {et~al.}}]{wu2019tomonaga}%
  \BibitemOpen
  \bibfield  {author} {\bibinfo {author} {\bibfnamefont {L.}~\bibnamefont
  {Wu}}, \bibinfo {author} {\bibfnamefont {S.}~\bibnamefont {Nikitin}},
  \bibinfo {author} {\bibfnamefont {Z.}~\bibnamefont {Wang}}, \bibinfo {author}
  {\bibfnamefont {W.}~\bibnamefont {Zhu}}, \bibinfo {author} {\bibfnamefont
  {C.~D.}\ \bibnamefont {Batista}}, \bibinfo {author} {\bibfnamefont
  {A.}~\bibnamefont {Tsvelik}}, \bibinfo {author} {\bibfnamefont {A.~M.}\
  \bibnamefont {Samarakoon}}, \bibinfo {author} {\bibfnamefont {D.~A.}\
  \bibnamefont {Tennant}}, \bibinfo {author} {\bibfnamefont {M.}~\bibnamefont
  {Brando}}, \bibinfo {author} {\bibfnamefont {L.}~\bibnamefont {Vasylechko}},
  \emph {et~al.},\ }\bibfield  {title} {\bibinfo {title} {Tomonaga--{L}uttinger
  liquid behavior and spinon confinement in {Y}b{A}l{O}$_{3}$},\ }\href@noop {}
  {\bibfield  {journal} {\bibinfo  {journal} {Nature communications}\ }\textbf
  {\bibinfo {volume} {10}},\ \bibinfo {pages} {698} (\bibinfo {year}
  {2019})}\BibitemShut {NoStop}%
\bibitem [{\citenamefont {Scheie}\ \emph {et~al.}(2023)\citenamefont {Scheie},
  \citenamefont {Ghioldi}, \citenamefont {Xing}, \citenamefont {Paddison},
  \citenamefont {Sherman}, \citenamefont {Dupont}, \citenamefont {Sanjeewa},
  \citenamefont {Lee}, \citenamefont {Woods}, \citenamefont {Abernathy} \emph
  {et~al.}}]{scheie2023proximate}%
  \BibitemOpen
  \bibfield  {author} {\bibinfo {author} {\bibfnamefont {A.}~\bibnamefont
  {Scheie}}, \bibinfo {author} {\bibfnamefont {E.}~\bibnamefont {Ghioldi}},
  \bibinfo {author} {\bibfnamefont {J.}~\bibnamefont {Xing}}, \bibinfo {author}
  {\bibfnamefont {J.}~\bibnamefont {Paddison}}, \bibinfo {author}
  {\bibfnamefont {N.}~\bibnamefont {Sherman}}, \bibinfo {author} {\bibfnamefont
  {M.}~\bibnamefont {Dupont}}, \bibinfo {author} {\bibfnamefont
  {L.}~\bibnamefont {Sanjeewa}}, \bibinfo {author} {\bibfnamefont
  {S.}~\bibnamefont {Lee}}, \bibinfo {author} {\bibfnamefont {A.}~\bibnamefont
  {Woods}}, \bibinfo {author} {\bibfnamefont {D.}~\bibnamefont {Abernathy}},
  \emph {et~al.},\ }\bibfield  {title} {\bibinfo {title} {Proximate spin liquid
  and fractionalization in the triangular antiferromagnet {KYbSe$_{2}$}},\
  }\href@noop {} {\bibfield  {journal} {\bibinfo  {journal} {Nature Physics}\
  ,\ \bibinfo {pages} {1}} (\bibinfo {year} {2023})}\BibitemShut {NoStop}%
\bibitem [{\citenamefont {Scheie}\ \emph {et~al.}(2022)\citenamefont {Scheie},
  \citenamefont {Kamiya}, \citenamefont {Zhang}, \citenamefont {Lee},
  \citenamefont {Woods}, \citenamefont {Gonzalez}, \citenamefont {Bernu},
  \citenamefont {Xing}, \citenamefont {Pajerowski}, \citenamefont {Zhou} \emph
  {et~al.}}]{scheie2022non}%
  \BibitemOpen
  \bibfield  {author} {\bibinfo {author} {\bibfnamefont {A.}~\bibnamefont
  {Scheie}}, \bibinfo {author} {\bibfnamefont {Y.}~\bibnamefont {Kamiya}},
  \bibinfo {author} {\bibfnamefont {H.}~\bibnamefont {Zhang}}, \bibinfo
  {author} {\bibfnamefont {S.}~\bibnamefont {Lee}}, \bibinfo {author}
  {\bibfnamefont {A.}~\bibnamefont {Woods}}, \bibinfo {author} {\bibfnamefont
  {M.}~\bibnamefont {Gonzalez}}, \bibinfo {author} {\bibfnamefont
  {B.}~\bibnamefont {Bernu}}, \bibinfo {author} {\bibfnamefont
  {J.}~\bibnamefont {Xing}}, \bibinfo {author} {\bibfnamefont {D.}~\bibnamefont
  {Pajerowski}}, \bibinfo {author} {\bibfnamefont {H.}~\bibnamefont {Zhou}},
  \emph {et~al.},\ }\bibfield  {title} {\bibinfo {title} {Non-linear magnons in
  the 1/3 magnetization plateau of a proximate quantum spin liquid},\
  }\href@noop {} {\bibfield  {journal} {\bibinfo  {journal} {arXiv preprint
  arXiv:2207.14785}\ } (\bibinfo {year} {2022})}\BibitemShut {NoStop}%
\bibitem [{\citenamefont {Xing}\ \emph {et~al.}(2021)\citenamefont {Xing},
  \citenamefont {Sanjeewa}, \citenamefont {May},\ and\ \citenamefont
  {Sefat}}]{xing2021synthesis}%
  \BibitemOpen
  \bibfield  {author} {\bibinfo {author} {\bibfnamefont {J.}~\bibnamefont
  {Xing}}, \bibinfo {author} {\bibfnamefont {L.~D.}\ \bibnamefont {Sanjeewa}},
  \bibinfo {author} {\bibfnamefont {A.~F.}\ \bibnamefont {May}},\ and\ \bibinfo
  {author} {\bibfnamefont {A.~S.}\ \bibnamefont {Sefat}},\ }\bibfield  {title}
  {\bibinfo {title} {Synthesis and anisotropic magnetism in quantum spin liquid
  candidates {$A$YbSe$_{2}$ (A= K and Rb)}},\ }\href@noop {} {\bibfield
  {journal} {\bibinfo  {journal} {APL Materials}\ }\textbf {\bibinfo {volume}
  {9}} (\bibinfo {year} {2021})}\BibitemShut {NoStop}%
\bibitem [{\citenamefont {Yamamoto}\ \emph {et~al.}(2014)\citenamefont
  {Yamamoto}, \citenamefont {Marmorini},\ and\ \citenamefont
  {Danshita}}]{yamamoto2014quantum}%
  \BibitemOpen
  \bibfield  {author} {\bibinfo {author} {\bibfnamefont {D.}~\bibnamefont
  {Yamamoto}}, \bibinfo {author} {\bibfnamefont {G.}~\bibnamefont
  {Marmorini}},\ and\ \bibinfo {author} {\bibfnamefont {I.}~\bibnamefont
  {Danshita}},\ }\bibfield  {title} {\bibinfo {title} {Quantum phase diagram of
  the triangular-lattice {$XXZ$} model in a magnetic field},\ }\href@noop {}
  {\bibfield  {journal} {\bibinfo  {journal} {Physical Review Letters}\
  }\textbf {\bibinfo {volume} {112}},\ \bibinfo {pages} {127203} (\bibinfo
  {year} {2014})}\BibitemShut {NoStop}%
\bibitem [{\citenamefont {Gopal}(1966)}]{Gopalsch}%
  \BibitemOpen
  \bibfield  {author} {\bibinfo {author} {\bibfnamefont {E.}~\bibnamefont
  {Gopal}},\ }\href@noop {} {\emph {\bibinfo {title} {Magnetic Contribution to
  Specific Heats}}}\ (\bibinfo  {publisher} {Springer},\ \bibinfo {address}
  {Boston},\ \bibinfo {year} {1966})\ pp.\ \bibinfo {pages}
  {102--111}\BibitemShut {NoStop}%
\bibitem [{\citenamefont {Bleaney}(1963)}]{Bleaney1963}%
  \BibitemOpen
  \bibfield  {author} {\bibinfo {author} {\bibfnamefont {B.}~\bibnamefont
  {Bleaney}},\ }\bibfield  {title} {\bibinfo {title} {Hyperfine interactions in
  rare-earth metals},\ }\href {https://doi.org/10.1063/1.1729355} {\bibfield
  {journal} {\bibinfo  {journal} {Journal of Applied Physics}\ }\textbf
  {\bibinfo {volume} {34}},\ \bibinfo {pages} {1024} (\bibinfo {year}
  {1963})}\BibitemShut {NoStop}%
\bibitem [{\citenamefont {Scheie}(2019)}]{scheie2019exotic}%
  \BibitemOpen
  \bibfield  {author} {\bibinfo {author} {\bibfnamefont {A.}~\bibnamefont
  {Scheie}},\ }\emph {\bibinfo {title} {Exotic Magnetism in Frustrated
  Pyrochlore-Based Magnets}},\ \href
  {https://jscholarship.library.jhu.edu/bitstream/handle/1774.2/62185/SCHEIE-DISSERTATION-2019.pdf?sequence=1}
  {Ph.D. thesis},\ \bibinfo  {school} {The Johns Hopkins University} (\bibinfo
  {year} {2019})\BibitemShut {NoStop}%
\bibitem [{\citenamefont {Ishii}\ \emph {et~al.}(2011)\citenamefont {Ishii},
  \citenamefont {Tanaka}, \citenamefont {Onuma}, \citenamefont {Nambu},
  \citenamefont {Tokunaga}, \citenamefont {Sakakibara}, \citenamefont
  {Kawashima}, \citenamefont {Maeno}, \citenamefont {Broholm}, \citenamefont
  {Gautreaux} \emph {et~al.}}]{ishii2011successive}%
  \BibitemOpen
  \bibfield  {author} {\bibinfo {author} {\bibfnamefont {R.}~\bibnamefont
  {Ishii}}, \bibinfo {author} {\bibfnamefont {S.}~\bibnamefont {Tanaka}},
  \bibinfo {author} {\bibfnamefont {K.}~\bibnamefont {Onuma}}, \bibinfo
  {author} {\bibfnamefont {Y.}~\bibnamefont {Nambu}}, \bibinfo {author}
  {\bibfnamefont {M.}~\bibnamefont {Tokunaga}}, \bibinfo {author}
  {\bibfnamefont {T.}~\bibnamefont {Sakakibara}}, \bibinfo {author}
  {\bibfnamefont {N.}~\bibnamefont {Kawashima}}, \bibinfo {author}
  {\bibfnamefont {Y.}~\bibnamefont {Maeno}}, \bibinfo {author} {\bibfnamefont
  {C.}~\bibnamefont {Broholm}}, \bibinfo {author} {\bibfnamefont {D.~P.}\
  \bibnamefont {Gautreaux}}, \emph {et~al.},\ }\bibfield  {title} {\bibinfo
  {title} {Successive phase transitions and phase diagrams for the
  quasi-two-dimensional easy-axis triangular antiferromagnet {Rb$_4$Mn(MoO$_4$)
  $_3$}},\ }\href@noop {} {\bibfield  {journal} {\bibinfo  {journal}
  {Europhysics Letters}\ }\textbf {\bibinfo {volume} {94}},\ \bibinfo {pages}
  {17001} (\bibinfo {year} {2011})}\BibitemShut {NoStop}%
\bibitem [{\citenamefont {Zapf}\ \emph {et~al.}(2014)\citenamefont {Zapf},
  \citenamefont {Jaime},\ and\ \citenamefont {Batista}}]{zapf2014bose}%
  \BibitemOpen
  \bibfield  {author} {\bibinfo {author} {\bibfnamefont {V.}~\bibnamefont
  {Zapf}}, \bibinfo {author} {\bibfnamefont {M.}~\bibnamefont {Jaime}},\ and\
  \bibinfo {author} {\bibfnamefont {C.}~\bibnamefont {Batista}},\ }\bibfield
  {title} {\bibinfo {title} {Bose-einstein condensation in quantum magnets},\
  }\href@noop {} {\bibfield  {journal} {\bibinfo  {journal} {Reviews of Modern
  Physics}\ }\textbf {\bibinfo {volume} {86}},\ \bibinfo {pages} {563}
  (\bibinfo {year} {2014})}\BibitemShut {NoStop}%
\bibitem [{\citenamefont {Lee}\ \emph {et~al.}(2023)\citenamefont {Lee},
  \citenamefont {Sch{\"o}nemann}, \citenamefont {Zhang}, \citenamefont
  {Dahlbom}, \citenamefont {Jang}, \citenamefont {Do}, \citenamefont
  {Christianson}, \citenamefont {Cheong}, \citenamefont {Park}, \citenamefont
  {Brosha} \emph {et~al.}}]{lee2023field}%
  \BibitemOpen
  \bibfield  {author} {\bibinfo {author} {\bibfnamefont {M.}~\bibnamefont
  {Lee}}, \bibinfo {author} {\bibfnamefont {R.}~\bibnamefont {Sch{\"o}nemann}},
  \bibinfo {author} {\bibfnamefont {H.}~\bibnamefont {Zhang}}, \bibinfo
  {author} {\bibfnamefont {D.}~\bibnamefont {Dahlbom}}, \bibinfo {author}
  {\bibfnamefont {T.-H.}\ \bibnamefont {Jang}}, \bibinfo {author}
  {\bibfnamefont {S.-H.}\ \bibnamefont {Do}}, \bibinfo {author} {\bibfnamefont
  {A.~D.}\ \bibnamefont {Christianson}}, \bibinfo {author} {\bibfnamefont
  {S.-W.}\ \bibnamefont {Cheong}}, \bibinfo {author} {\bibfnamefont {J.-H.}\
  \bibnamefont {Park}}, \bibinfo {author} {\bibfnamefont {E.}~\bibnamefont
  {Brosha}}, \emph {et~al.},\ }\bibfield  {title} {\bibinfo {title}
  {Field-induced spin level crossings within a quasi-x y antiferromagnetic
  state in {Ba$_2$FeSi$_2$O$_7$}},\ }\href@noop {} {\bibfield  {journal}
  {\bibinfo  {journal} {Physical Review B}\ }\textbf {\bibinfo {volume}
  {107}},\ \bibinfo {pages} {144427} (\bibinfo {year} {2023})}\BibitemShut
  {NoStop}%
\bibitem [{\citenamefont {Farnell}\ \emph {et~al.}(2009)\citenamefont
  {Farnell}, \citenamefont {Zinke}, \citenamefont {Schulenburg},\ and\
  \citenamefont {Richter}}]{farnell2009high}%
  \BibitemOpen
  \bibfield  {author} {\bibinfo {author} {\bibfnamefont {D.~J.}\ \bibnamefont
  {Farnell}}, \bibinfo {author} {\bibfnamefont {R.}~\bibnamefont {Zinke}},
  \bibinfo {author} {\bibfnamefont {J.}~\bibnamefont {Schulenburg}},\ and\
  \bibinfo {author} {\bibfnamefont {J.}~\bibnamefont {Richter}},\ }\bibfield
  {title} {\bibinfo {title} {High-order coupled cluster method study of
  frustrated and unfrustrated quantum magnets in external magnetic fields},\
  }\href@noop {} {\bibfield  {journal} {\bibinfo  {journal} {Journal of
  Physics: Condensed Matter}\ }\textbf {\bibinfo {volume} {21}},\ \bibinfo
  {pages} {406002} (\bibinfo {year} {2009})}\BibitemShut {NoStop}%
\bibitem [{\citenamefont {Jolicoeur}\ \emph {et~al.}(1990)\citenamefont
  {Jolicoeur}, \citenamefont {Dagotto}, \citenamefont {Gagliano},\ and\
  \citenamefont {Bacci}}]{Jolicoeur1990tri}%
  \BibitemOpen
  \bibfield  {author} {\bibinfo {author} {\bibfnamefont {T.}~\bibnamefont
  {Jolicoeur}}, \bibinfo {author} {\bibfnamefont {E.}~\bibnamefont {Dagotto}},
  \bibinfo {author} {\bibfnamefont {E.}~\bibnamefont {Gagliano}},\ and\
  \bibinfo {author} {\bibfnamefont {S.}~\bibnamefont {Bacci}},\ }\bibfield
  {title} {\bibinfo {title} {Ground-state properties of the s=1/2 heisenberg
  antiferromagnet on a triangular lattice},\ }\href@noop {} {\bibfield
  {journal} {\bibinfo  {journal} {Physical Review B}\ }\textbf {\bibinfo
  {volume} {42}},\ \bibinfo {pages} {4800} (\bibinfo {year}
  {1990})}\BibitemShut {NoStop}%
\bibitem [{\citenamefont {Ran}\ \emph {et~al.}(2007)\citenamefont {Ran},
  \citenamefont {Hermele}, \citenamefont {Lee},\ and\ \citenamefont
  {Wen}}]{Ran2007linear}%
  \BibitemOpen
  \bibfield  {author} {\bibinfo {author} {\bibfnamefont {Y.}~\bibnamefont
  {Ran}}, \bibinfo {author} {\bibfnamefont {M.}~\bibnamefont {Hermele}},
  \bibinfo {author} {\bibfnamefont {P.}~\bibnamefont {Lee}},\ and\ \bibinfo
  {author} {\bibfnamefont {X.-G.}\ \bibnamefont {Wen}},\ }\bibfield  {title}
  {\bibinfo {title} {Projected-wave-function study of the spin-1/2 heisenberg
  model on the kagome lattice},\ }\href@noop {} {\bibfield  {journal} {\bibinfo
   {journal} {Physical Review Letters}\ }\textbf {\bibinfo {volume} {98}},\
  \bibinfo {pages} {117205} (\bibinfo {year} {2007})}\BibitemShut {NoStop}%
\bibitem [{\citenamefont {Barth{\'e}lemy}\ \emph {et~al.}(2022)\citenamefont
  {Barth{\'e}lemy}, \citenamefont {Demuer}, \citenamefont {Marcenat},
  \citenamefont {Klein}, \citenamefont {Bernu}, \citenamefont {Messio},
  \citenamefont {Vel{\'a}zquez}, \citenamefont {Kermarrec}, \citenamefont
  {Bert},\ and\ \citenamefont {Mendels}}]{Barthelemy2022linear}%
  \BibitemOpen
  \bibfield  {author} {\bibinfo {author} {\bibfnamefont {Q.}~\bibnamefont
  {Barth{\'e}lemy}}, \bibinfo {author} {\bibfnamefont {A.}~\bibnamefont
  {Demuer}}, \bibinfo {author} {\bibfnamefont {C.}~\bibnamefont {Marcenat}},
  \bibinfo {author} {\bibfnamefont {T.}~\bibnamefont {Klein}}, \bibinfo
  {author} {\bibfnamefont {B.}~\bibnamefont {Bernu}}, \bibinfo {author}
  {\bibfnamefont {L.}~\bibnamefont {Messio}}, \bibinfo {author} {\bibfnamefont
  {M.}~\bibnamefont {Vel{\'a}zquez}}, \bibinfo {author} {\bibfnamefont
  {E.}~\bibnamefont {Kermarrec}}, \bibinfo {author} {\bibfnamefont
  {F.}~\bibnamefont {Bert}},\ and\ \bibinfo {author} {\bibfnamefont
  {P.}~\bibnamefont {Mendels}},\ }\bibfield  {title} {\bibinfo {title}
  {Specific heat of the kagome antiferromagnet herbertsmithite in high magnetic
  fields},\ }\href@noop {} {\bibfield  {journal} {\bibinfo  {journal} {Physical
  Review X}\ }\textbf {\bibinfo {volume} {12}},\ \bibinfo {pages} {011014}
  (\bibinfo {year} {2022})}\BibitemShut {NoStop}%
\bibitem [{\citenamefont {Liu}\ \emph {et~al.}(2021)\citenamefont {Liu},
  \citenamefont {Liu}, \citenamefont {Yuan}, \citenamefont {Li}, \citenamefont
  {Xie}, \citenamefont {Chen}, \citenamefont {Zhang}, \citenamefont {Xu},
  \citenamefont {Tong}, \citenamefont {Wang},\ and\ \citenamefont
  {Li}}]{Liu2021linear}%
  \BibitemOpen
  \bibfield  {author} {\bibinfo {author} {\bibfnamefont {J.}~\bibnamefont
  {Liu}}, \bibinfo {author} {\bibfnamefont {B.}~\bibnamefont {Liu}}, \bibinfo
  {author} {\bibfnamefont {L.}~\bibnamefont {Yuan}}, \bibinfo {author}
  {\bibfnamefont {B.}~\bibnamefont {Li}}, \bibinfo {author} {\bibfnamefont
  {L.}~\bibnamefont {Xie}}, \bibinfo {author} {\bibfnamefont {X.}~\bibnamefont
  {Chen}}, \bibinfo {author} {\bibfnamefont {H.}~\bibnamefont {Zhang}},
  \bibinfo {author} {\bibfnamefont {D.}~\bibnamefont {Xu}}, \bibinfo {author}
  {\bibfnamefont {W.}~\bibnamefont {Tong}}, \bibinfo {author} {\bibfnamefont
  {J.}~\bibnamefont {Wang}},\ and\ \bibinfo {author} {\bibfnamefont
  {Y.}~\bibnamefont {Li}},\ }\bibfield  {title} {\bibinfo {title} {Frustrated
  magnetism of the triangular-lattice antiferromagnets {$\alpha$-CrOOH} and
  {$\alpha$-CrOOD}},\ }\href@noop {} {\bibfield  {journal} {\bibinfo  {journal}
  {New Journal of Physics}\ }\textbf {\bibinfo {volume} {23}},\ \bibinfo
  {pages} {033040} (\bibinfo {year} {2021})}\BibitemShut {NoStop}%
\bibitem [{\citenamefont {Sindzingre}\ \emph {et~al.}(2000)\citenamefont
  {Sindzingre}, \citenamefont {Misguich}, \citenamefont {Lhuilier},
  \citenamefont {Bernu}, \citenamefont {Pierre}, \citenamefont {Waldtmann},\
  and\ \citenamefont {Everts}}]{Sindzingre2000linear}%
  \BibitemOpen
  \bibfield  {author} {\bibinfo {author} {\bibfnamefont {P.}~\bibnamefont
  {Sindzingre}}, \bibinfo {author} {\bibfnamefont {G.}~\bibnamefont
  {Misguich}}, \bibinfo {author} {\bibfnamefont {C.}~\bibnamefont {Lhuilier}},
  \bibinfo {author} {\bibfnamefont {B.}~\bibnamefont {Bernu}}, \bibinfo
  {author} {\bibfnamefont {L.}~\bibnamefont {Pierre}}, \bibinfo {author}
  {\bibfnamefont {C.}~\bibnamefont {Waldtmann}},\ and\ \bibinfo {author}
  {\bibfnamefont {H.-U.}\ \bibnamefont {Everts}},\ }\bibfield  {title}
  {\bibinfo {title} {Magnetothermodynamics of the spin- 1/2 kagome
  antiferromagnet},\ }\href@noop {} {\bibfield  {journal} {\bibinfo  {journal}
  {Physical Review Letters}\ }\textbf {\bibinfo {volume} {84}},\ \bibinfo
  {pages} {13} (\bibinfo {year} {2000})}\BibitemShut {NoStop}%
\bibitem [{\citenamefont {Samulon}\ \emph {et~al.}(2008)\citenamefont
  {Samulon}, \citenamefont {Jo}, \citenamefont {Sengupta}, \citenamefont
  {Batista}, \citenamefont {Jaime}, \citenamefont {Balicas},\ and\
  \citenamefont {Fisher}}]{samulon2008ordered}%
  \BibitemOpen
  \bibfield  {author} {\bibinfo {author} {\bibfnamefont {E.}~\bibnamefont
  {Samulon}}, \bibinfo {author} {\bibfnamefont {Y.-J.}\ \bibnamefont {Jo}},
  \bibinfo {author} {\bibfnamefont {P.}~\bibnamefont {Sengupta}}, \bibinfo
  {author} {\bibfnamefont {C.}~\bibnamefont {Batista}}, \bibinfo {author}
  {\bibfnamefont {M.}~\bibnamefont {Jaime}}, \bibinfo {author} {\bibfnamefont
  {L.}~\bibnamefont {Balicas}},\ and\ \bibinfo {author} {\bibfnamefont
  {I.}~\bibnamefont {Fisher}},\ }\bibfield  {title} {\bibinfo {title} {Ordered
  magnetic phases of the frustrated spin-dimer compound ba 3 mn 2 o 8},\
  }\href@noop {} {\bibfield  {journal} {\bibinfo  {journal} {Physical Review
  B}\ }\textbf {\bibinfo {volume} {77}},\ \bibinfo {pages} {214441} (\bibinfo
  {year} {2008})}\BibitemShut {NoStop}%
\bibitem [{\citenamefont {Chen}\ \emph {et~al.}(2013)\citenamefont {Chen},
  \citenamefont {Ju}, \citenamefont {Jiang}, \citenamefont {Starykh},\ and\
  \citenamefont {Balents}}]{chen2013ground}%
  \BibitemOpen
  \bibfield  {author} {\bibinfo {author} {\bibfnamefont {R.}~\bibnamefont
  {Chen}}, \bibinfo {author} {\bibfnamefont {H.}~\bibnamefont {Ju}}, \bibinfo
  {author} {\bibfnamefont {H.-C.}\ \bibnamefont {Jiang}}, \bibinfo {author}
  {\bibfnamefont {O.~A.}\ \bibnamefont {Starykh}},\ and\ \bibinfo {author}
  {\bibfnamefont {L.}~\bibnamefont {Balents}},\ }\bibfield  {title} {\bibinfo
  {title} {Ground states of spin-1 2 triangular antiferromagnets in a magnetic
  field},\ }\href@noop {} {\bibfield  {journal} {\bibinfo  {journal} {Physical
  Review B}\ }\textbf {\bibinfo {volume} {87}},\ \bibinfo {pages} {165123}
  (\bibinfo {year} {2013})}\BibitemShut {NoStop}%
\bibitem [{\citenamefont {Syromyatnikov}(2023)}]{syromyatnikov2023unusual}%
  \BibitemOpen
  \bibfield  {author} {\bibinfo {author} {\bibfnamefont {A.}~\bibnamefont
  {Syromyatnikov}},\ }\bibfield  {title} {\bibinfo {title} {Unusual dynamics of
  spin-12 antiferromagnets on the triangular lattice in magnetic field},\
  }\href@noop {} {\bibfield  {journal} {\bibinfo  {journal} {Annals of
  Physics}\ }\textbf {\bibinfo {volume} {454}},\ \bibinfo {pages} {169342}
  (\bibinfo {year} {2023})}\BibitemShut {NoStop}%
\bibitem [{\citenamefont {Honecker}\ \emph {et~al.}(2004)\citenamefont
  {Honecker}, \citenamefont {Schulenburg},\ and\ \citenamefont
  {Richter}}]{honecker2004magnetization}%
  \BibitemOpen
  \bibfield  {author} {\bibinfo {author} {\bibfnamefont {A.}~\bibnamefont
  {Honecker}}, \bibinfo {author} {\bibfnamefont {J.}~\bibnamefont
  {Schulenburg}},\ and\ \bibinfo {author} {\bibfnamefont {J.}~\bibnamefont
  {Richter}},\ }\bibfield  {title} {\bibinfo {title} {Magnetization plateaus in
  frustrated antiferromagnetic quantum spin models},\ }\href@noop {} {\bibfield
   {journal} {\bibinfo  {journal} {Journal of Physics: Condensed Matter}\
  }\textbf {\bibinfo {volume} {16}},\ \bibinfo {pages} {S749} (\bibinfo {year}
  {2004})}\BibitemShut {NoStop}%
\bibitem [{\citenamefont {Tsujii}\ \emph {et~al.}(2007)\citenamefont {Tsujii},
  \citenamefont {Rotundu}, \citenamefont {Ono}, \citenamefont {Tanaka},
  \citenamefont {Andraka}, \citenamefont {Ingersent},\ and\ \citenamefont
  {Takano}}]{Tsujii2007CCB}%
  \BibitemOpen
  \bibfield  {author} {\bibinfo {author} {\bibfnamefont {H.}~\bibnamefont
  {Tsujii}}, \bibinfo {author} {\bibfnamefont {C.}~\bibnamefont {Rotundu}},
  \bibinfo {author} {\bibfnamefont {T.}~\bibnamefont {Ono}}, \bibinfo {author}
  {\bibfnamefont {H.}~\bibnamefont {Tanaka}}, \bibinfo {author} {\bibfnamefont
  {B.}~\bibnamefont {Andraka}}, \bibinfo {author} {\bibfnamefont
  {K.}~\bibnamefont {Ingersent}},\ and\ \bibinfo {author} {\bibfnamefont
  {Y.}~\bibnamefont {Takano}},\ }\bibfield  {title} {\bibinfo {title}
  {Thermodynamics of the up-up-down phase of the s=1/2 triangular-lattice
  antiferromagnet {Cs$_2$CuBr$_4$}},\ }\href@noop {} {\bibfield  {journal}
  {\bibinfo  {journal} {Physical Review B}\ }\textbf {\bibinfo {volume} {76}},\
  \bibinfo {pages} {060406(R)} (\bibinfo {year} {2007})}\BibitemShut {NoStop}%
\bibitem [{\citenamefont {Hwang}\ \emph {et~al.}(2012)\citenamefont {Hwang},
  \citenamefont {Choi}, \citenamefont {Ye}, \citenamefont {Dela~Cruz},
  \citenamefont {Xin}, \citenamefont {Zhou},\ and\ \citenamefont
  {Schlottmann}}]{Hwang2012uud}%
  \BibitemOpen
  \bibfield  {author} {\bibinfo {author} {\bibfnamefont {J.}~\bibnamefont
  {Hwang}}, \bibinfo {author} {\bibfnamefont {E.}~\bibnamefont {Choi}},
  \bibinfo {author} {\bibfnamefont {F.}~\bibnamefont {Ye}}, \bibinfo {author}
  {\bibfnamefont {C.}~\bibnamefont {Dela~Cruz}}, \bibinfo {author}
  {\bibfnamefont {Y.}~\bibnamefont {Xin}}, \bibinfo {author} {\bibfnamefont
  {H.}~\bibnamefont {Zhou}},\ and\ \bibinfo {author} {\bibfnamefont
  {P.}~\bibnamefont {Schlottmann}},\ }\bibfield  {title} {\bibinfo {title}
  {Successive magnetic phase transitions and multiferroicity in the spin-one
  triangular-lattice antiferromagnet {Ba$_3$NiNb$_2$O$_9$}},\ }\href@noop {}
  {\bibfield  {journal} {\bibinfo  {journal} {Physical Review Letters}\
  }\textbf {\bibinfo {volume} {109}},\ \bibinfo {pages} {257205} (\bibinfo
  {year} {2012})}\BibitemShut {NoStop}%
\bibitem [{\citenamefont {Lee}\ \emph {et~al.}(2017)\citenamefont {Lee},
  \citenamefont {Choi}, \citenamefont {Ma}, \citenamefont {Sinclair},
  \citenamefont {Dela~Cruz},\ and\ \citenamefont {Zhou}}]{Lee2017uud}%
  \BibitemOpen
  \bibfield  {author} {\bibinfo {author} {\bibfnamefont {M.}~\bibnamefont
  {Lee}}, \bibinfo {author} {\bibfnamefont {E.}~\bibnamefont {Choi}}, \bibinfo
  {author} {\bibfnamefont {J.}~\bibnamefont {Ma}}, \bibinfo {author}
  {\bibfnamefont {R.}~\bibnamefont {Sinclair}}, \bibinfo {author}
  {\bibfnamefont {C.}~\bibnamefont {Dela~Cruz}},\ and\ \bibinfo {author}
  {\bibfnamefont {H.}~\bibnamefont {Zhou}},\ }\bibfield  {title} {\bibinfo
  {title} {Magnetic and electric properties of triangular lattice
  antiferromagnets {Ba$_3$ATa$_2$O$_9$} ({A} = ni and co)},\ }\href@noop {}
  {\bibfield  {journal} {\bibinfo  {journal} {Materials Research Bulletin}\
  }\textbf {\bibinfo {volume} {88}},\ \bibinfo {pages} {308} (\bibinfo {year}
  {2017})}\BibitemShut {NoStop}%
\bibitem [{\citenamefont {Gao}\ \emph {et~al.}(2022)\citenamefont {Gao},
  \citenamefont {Fan}, \citenamefont {Li}, \citenamefont {Yang}, \citenamefont
  {Zeng}, \citenamefont {Sheng}, \citenamefont {Zhong}, \citenamefont {Qi},
  \citenamefont {Wan},\ and\ \citenamefont {Li}}]{Gao2022uud}%
  \BibitemOpen
  \bibfield  {author} {\bibinfo {author} {\bibfnamefont {Y.}~\bibnamefont
  {Gao}}, \bibinfo {author} {\bibfnamefont {Y.-C.}\ \bibnamefont {Fan}},
  \bibinfo {author} {\bibfnamefont {H.}~\bibnamefont {Li}}, \bibinfo {author}
  {\bibfnamefont {F.}~\bibnamefont {Yang}}, \bibinfo {author} {\bibfnamefont
  {X.-T.}\ \bibnamefont {Zeng}}, \bibinfo {author} {\bibfnamefont {X.-T.}\
  \bibnamefont {Sheng}}, \bibinfo {author} {\bibfnamefont {R.}~\bibnamefont
  {Zhong}}, \bibinfo {author} {\bibfnamefont {Y.}~\bibnamefont {Qi}}, \bibinfo
  {author} {\bibfnamefont {Y.}~\bibnamefont {Wan}},\ and\ \bibinfo {author}
  {\bibfnamefont {W.}~\bibnamefont {Li}},\ }\bibfield  {title} {\bibinfo
  {title} {Spin supersolidity in nearly ideal easy-axis triangular quantum
  antiferromagnet {Na$_2$BaCo(PO$_4$)$_2$}},\ }\href@noop {} {\bibfield
  {journal} {\bibinfo  {journal} {npj Quantum Materials}\ }\textbf {\bibinfo
  {volume} {7}},\ \bibinfo {pages} {89} (\bibinfo {year} {2022})}\BibitemShut
  {NoStop}%
\bibitem [{\citenamefont {Smirnov}\ \emph {et~al.}(2007)\citenamefont
  {Smirnov}, \citenamefont {Yashiro}, \citenamefont {Kimura}, \citenamefont
  {Hagiwara}, \citenamefont {Narumi}, \citenamefont {Kindo}, \citenamefont
  {Kikkawa}, \citenamefont {Katsumata}, \citenamefont {Shapiro},\ and\
  \citenamefont {Demianets}}]{Smirnov2007uud}%
  \BibitemOpen
  \bibfield  {author} {\bibinfo {author} {\bibfnamefont {A.}~\bibnamefont
  {Smirnov}}, \bibinfo {author} {\bibfnamefont {H.}~\bibnamefont {Yashiro}},
  \bibinfo {author} {\bibfnamefont {S.}~\bibnamefont {Kimura}}, \bibinfo
  {author} {\bibfnamefont {M.}~\bibnamefont {Hagiwara}}, \bibinfo {author}
  {\bibfnamefont {Y.}~\bibnamefont {Narumi}}, \bibinfo {author} {\bibfnamefont
  {K.}~\bibnamefont {Kindo}}, \bibinfo {author} {\bibfnamefont
  {A.}~\bibnamefont {Kikkawa}}, \bibinfo {author} {\bibfnamefont
  {K.}~\bibnamefont {Katsumata}}, \bibinfo {author} {\bibfnamefont
  {A.}~\bibnamefont {Shapiro}},\ and\ \bibinfo {author} {\bibfnamefont
  {L.}~\bibnamefont {Demianets}},\ }\bibfield  {title} {\bibinfo {title}
  {Triangular lattice antiferromagnet {RbFe(MoO$_4$)$_2$} in high magnetic
  fields},\ }\href@noop {} {\bibfield  {journal} {\bibinfo  {journal} {Physical
  Review B}\ }\textbf {\bibinfo {volume} {75}},\ \bibinfo {pages} {134412}
  (\bibinfo {year} {2007})}\BibitemShut {NoStop}%
\bibitem [{\citenamefont {Zapf}\ and\ \citenamefont
  {Jaime}(2014)}]{Viv2014mce}%
  \BibitemOpen
  \bibfield  {author} {\bibinfo {author} {\bibfnamefont {V.}~\bibnamefont
  {Zapf}}\ and\ \bibinfo {author} {\bibfnamefont {M.}~\bibnamefont {Jaime}},\
  }\bibfield  {title} {\bibinfo {title} {Bose-einstein condensation in quantum
  magnets},\ }\href@noop {} {\bibfield  {journal} {\bibinfo  {journal} {Reviews
  of Modern Physics}\ }\textbf {\bibinfo {volume} {86}},\ \bibinfo {pages}
  {563} (\bibinfo {year} {2014})}\BibitemShut {NoStop}%
\bibitem [{\citenamefont {Seabra}\ \emph {et~al.}(2016)\citenamefont {Seabra},
  \citenamefont {Sindzingre}, \citenamefont {Momoi},\ and\ \citenamefont
  {Shannon}}]{Seabra2016square}%
  \BibitemOpen
  \bibfield  {author} {\bibinfo {author} {\bibfnamefont {L.}~\bibnamefont
  {Seabra}}, \bibinfo {author} {\bibfnamefont {P.}~\bibnamefont {Sindzingre}},
  \bibinfo {author} {\bibfnamefont {T.}~\bibnamefont {Momoi}},\ and\ \bibinfo
  {author} {\bibfnamefont {N.}~\bibnamefont {Shannon}},\ }\bibfield  {title}
  {\bibinfo {title} {Novel phases in a square-lattice frustrated ferromagnet;
  1/3 -magnetization plateau, helicoidal spin liquid, and vortex crystal},\
  }\href@noop {} {\bibfield  {journal} {\bibinfo  {journal} {Physical Review
  B}\ }\textbf {\bibinfo {volume} {93}},\ \bibinfo {pages} {085132} (\bibinfo
  {year} {2016})}\BibitemShut {NoStop}%
\bibitem [{\citenamefont {Pankratova}\ \emph {et~al.}(2021)\citenamefont
  {Pankratova}, \citenamefont {Igoshev},\ and\ \citenamefont
  {Irkhin}}]{Pankratova2021cubic}%
  \BibitemOpen
  \bibfield  {author} {\bibinfo {author} {\bibfnamefont {A.}~\bibnamefont
  {Pankratova}}, \bibinfo {author} {\bibfnamefont {P.}~\bibnamefont
  {Igoshev}},\ and\ \bibinfo {author} {\bibfnamefont {V.}~\bibnamefont
  {Irkhin}},\ }\bibfield  {title} {\bibinfo {title} {Incommensurate magnetic
  order in rare earth and transition metal compounds with local moments},\
  }\href@noop {} {\bibfield  {journal} {\bibinfo  {journal} {Journal of
  Physics: Condensed Matter}\ }\textbf {\bibinfo {volume} {33}},\ \bibinfo
  {pages} {375802} (\bibinfo {year} {2021})}\BibitemShut {NoStop}%
\end{thebibliography}%

\end{document}